\documentclass[aps,twocolumn,floatfix,groupedaddress,nofootinbib]{revtex4}
\usepackage{csquotes}
\usepackage{graphicx,color}
\usepackage{float}
\usepackage[caption=false]{subfig}
\usepackage{amsmath}
\usepackage{amsmath}
\usepackage{multirow}
\usepackage{dsfont}
\usepackage[shortlabels]{enumitem}
\usepackage{epstopdf}

\begin{document}

\title{Bridge functions of classical one-component plasmas}
\author{\vspace*{-3.40mm}F. Lucco Castello and P. Tolias}
\affiliation{Space and Plasma Physics, Royal Institute of Technology, Stockholm, SE-100 44, Sweden\vspace*{-2.60mm}}
\begin{abstract}
\noindent In a recent paper, Lucco Castello \emph{et al.} [arXiv:2107.03537] performed systematic extractions of classical one-component plasma bridge functions from molecular dynamics simulations and provided an accurate parametrization that was incorporated in the novel isomorph-based empirically modified hypernetted chain approach for Yukawa one-component plasmas. Here, the extraction technique and parametrization strategy are described in detail, while the deficiencies of earlier efforts are discussed. The structural and thermodynamic predictions of the updated version of the integral equation theory approach are compared with extensive available simulation results revealing a truly unprecedented level of accuracy in the entire dense liquid region of the Yukawa phase diagram.
\end{abstract}
\maketitle

\section{Introduction}\label{sec:intro}

\noindent The physics of liquid state many-body systems that are composed of charged particles has evolved to a significant area of modern statistical mechanics. Such systems include ionic liquids and electrolyte solutions, dense astrophysical matter and non-ideal plasmas as well as complex (dusty) plasmas and colloidal suspensions\,\cite{Gaborboo,OCPrevBH}. Strongly coupled Coulomb liquids are characterized by a ratio of the average potential energy to the thermal energy that exceeds or is of the order of unity. The simplest model of Coulomb interacting liquids is the spatially homogeneous classical one-component plasma that consists of identical classical point charges which are embedded in a uniform inert background that guarantees that the overall charge neutrality is preserved\,\cite{OCPrevBH,IchiRev1}. In spite of the fact that this represents a substantial idealization, the one-component plasma has offered great insights to the behavior of naturally occurring systems and has been widely lauded as one of the most important model systems in physics\,\cite{Gaborboo,OCPrevBH,IchiRev1}.

One of the most fundamental problems in statistical mechanics of liquids concerns the accurate computation of thermodynamic properties and spatial density-density correlations with the sole knowledge of the interaction potential without resorting to computer simulations. The integral equation theory of liquids, that elegantly emerges from density expansions in the grand canonical ensemble with the aid of functional analysis and diagrammatic analysis\,\cite{Hansenbo,hbondbok,liquidbo,theorybo}, has been justifiably deduced to be the most appropriate framework to attack this problem\,\cite{perturba}. It consists of two formally exact equations that contain three unknown functions; an additional equation that describes the so-called bridge function is required for self-consistent closure of this formalism\,\cite{Hansenbo,hbondbok,liquidbo,theorybo}. However, the formal diagrammatic expansion of the bridge function, which is a very complicated functional of the interaction potential, has proven to converge very slowly\,\cite{Hansenbo,theorybo}; a rather anticipated fact given the impossibility of formally closing any many body problem in the absence of small parameters. Consequently, a large number of approximation schemes has been developed, whose effectiveness varies depending mainly on the potential softness and can only be reliably evaluated a posteriori through a systematic comparison with \enquote{exact} simulation results\,\cite{Hansenbo,BomontRe}.

The computation of bridge functions with structural input from computer simulations presents the only viable alternative to the seemingly endless parade of approximations of integral equation theory. In spite of the rather demanding design of the bridge function extraction simulations, the heavy computational cost and the necessity for meticulous uncertainty propagation analysis, \enquote{exact} bridge functions have been computed with structural input from Monte Carlo or molecular dynamics (MD) simulations for numerous liquid systems which include established model interactions such as hard spheres\,\cite{bridgHSO,bridgHS0,bridgHS1,bridgHS2} and their binary mixtures\,\cite{bridgHS0,bridgHS2}, Lennard-Jones interactions\,\cite{bridgLJ1,bridgLJ2,bridgLJ3,bridgLJ4}, inverse power law interactions\,\cite{bridIPL1}, one-component plasmas\,\cite{PollOCPB,bridOCP1,bridOCP2} and Yukawa one-component plasmas\,\cite{Yukawiso} as well as more realistic model interactions such as hard spheroids\,\cite{bridgHS3}, liquid metal interionic potentials\,\cite{bridgme1}, model 2-2 electrolytes\,\cite{bridgele}, molten salts\,\cite{bridgsal}, Lennard-Jones dipolar fluids\,\cite{bridgSTO} and even the extended simple point charge (SPC/E) site-site model of water\,\cite{bridgwat}. Unfortunately, the state-of-affairs is quite discouraging, since many bridge function extraction investigations concern the intermediate and long range exclusively, contain uncontrolled uncertainties and are restricted to a single or to few thermodynamic states. In fact, bridge function parametrizations are only available for hard spheres (intermediate \& long range)\cite{bridgHS4}, soft spheres (full range)\,\cite{bridIPL1} and one-component plasmas (entire range)\,\cite{bridOCP1,bridOCP2}. The bridge functions of hard spheres, one-component plasmas and Lennard-Jones systems are the most well-studied; indicative of the importance of the respective interactions. In particular, the one-component plasma bridge functions were amongst the first to be extracted\,\cite{PollOCPB} and to be parameterized\,\cite{bridOCP1}. Nevertheless, there are persistent problems with the available extractions and parametrizations.

In a recent Letter\,\cite{unpubli1}, we computed the bridge functions of classical one-component plasmas at $17$ thermodynamic states, spanning the whole dense liquid region, by utilizing radial distribution functions extracted from accurate standard canonical MD simulations in combination with cavity distribution functions extracted from long specially designed canonical MD simulations featuring tagged particle pairs. With this input, we constructed a very accurate closed-form bridge function parametrization that covers the full non-trivial range. This analytic description was embedded into the isomorph-based empirically modified hypernetted chain approximation\,\cite{ourwork1} that was then applied to Yukawa one-component plasmas leading to an unprecedent agreement with simulations.

In the present work, the devised bridge function extraction scheme is presented in full detail from the design of simulations to the propagation of uncertainties. Emphasis is put on peculiarities that stem from the long range nature of Coulomb interactions and are not present in other liquids. The adopted bridge function parametrization strategy is expanded on and the deficiencies of existing bridge function extractions \& parametrizations are discussed in depth. Moreover, the updated version of the isomorph-based empirically modified hypernetted chain approximation is presented and numerically solved in the entire dense liquid region of the Yukawa phase diagram. A systematic comparison is carried out with the original version and with available simulations in terms of radial distribution functions and excess internal energies.

\section{Integral equation theory and bridge function extraction methods}\label{sec:extraction_methods}

\noindent In case of a one-component pair-interacting isotropic system, the integral equation theory of liquids comprises the Ornstein-Zernike (OZ) integral equation and the formally exact non-linear closure condition\,\cite{Hansenbo,hbondbok,liquidbo,theorybo,BomontRe}, which read as
\begin{align}
h(r)&=c(r)+n\int c(r')h(|\boldsymbol{r}-\boldsymbol{r}'|)d^3r'\,,\label{OZequation}\\
g(r)&=\exp\left[-\beta u(r)+h(r)-c(r)+B(r)\right]\,,\label{OZclosure}
\end{align}
with $g(r)$ the radial distribution function, $h(r)=g(r)-1$ the total correlation function, $c(r)$ the direct correlation function, $B(r)$ the bridge function. Some auxiliary static two-particle correlation functions of importance are; the static structure factor $S(k)=1+n\widetilde{H}(k)$ where $\widetilde{H}(k)$ denotes the Fourier transform of the total correlation function, the indirect correlation function $\gamma(r)=h(r)-c(r)$, the mean force potential $\beta{w}(r)=-\ln{[g(r)]}$, the screening potential $\beta{H}(r)=\beta{u(r)}-\beta{w}(r)$ and the cavity distribution function $y(r)=g(r)\exp{[\beta{u}(r)]}$. It is evident that the screening potential is equal to the logarithm of the cavity distribution function, $\beta{H}(r)=\ln{[y(r)]}$, thus, in what follows, these terms will be used indiscriminately.

The integral equation theory set of equations requires a formally exact expression for the bridge function in order to be closed in a self-consistent manner\,\cite{Hansenbo,hbondbok,liquidbo,theorybo}. Two equivalent virial-type series representations have been derived within the diagrammatic analysis framework, where the bridge functions are graphically represented by highly connected diagrams, which contain neither nodal points nor articulation points and their root points do not form articulation pairs\,\cite{Hansenbo,theorybo}. More specifically, an exact $B[u]$ functional relation is available, known as $f-$bond expansion owing to the involvement of Mayer functions $f(r)=\exp{[-\beta{u}(r)]}-1$, that reads as $B[f]=\sum_{i=2}^{\infty}d_i(r;T)n^{i}$ where the coefficients $d_i(r;T)$ are given by a number of multi-dimensional integrals whose kernels are products of Mayer functions. In addition, the operation of topological reduction leads to an exact $B[h]$ functional relation, known as $h-$bond expansion owing to the involvement of total correlation functions, that reads as $B[h]=\sum_{i=2}^{\infty}b_i(r;n,T)n^{i}$ where the coefficients $b_i(r;n,T)$ are given by a number of multi-dimensional integrals whose kernels are now products of total correlation functions\,\cite{hbondbok}. In spite of numerous commendable efforts\,\cite{bridgev1,bridgev2,bridgev3,bridgev4}, it has proven to be rather hopeless to compute bridge functions through their definition, since both the original and the resummation series converge very slowly even at moderate densities, since the contributing diagram number exponentially increases with the density order and since the integrand complexity as well as integration order steadily increase with the density order. As a consequence, most theoretical attempts have either focused on the formulation of phenomenological closures which approximate the exact non-local functional of the total correlation function $B[h]$ with a local function of the indirect correlation function $B(h-c)$, or focused on optimizing thermodynamic state mappings to a reference system based on the approximate quasi-universality of bridge functions\,\cite{BomontRe}.

The bridge function is the collective term employed for highly connected irreducible diagrams and, thus, constitutes a rather abstract object of diagrammatic analysis. Unsurprisingly, it does not possess a microscopic representation in terms of the ensemble average of a function that depends on the instantaneous particle positions and it does not possess a physical interpretation in terms of a probability density. These imply that bridge functions cannot be directly extracted from computer simulations. Nevertheless, bridge functions can be indirectly extracted by exploiting the fact that radial distribution and cavity distribution functions can be directly extracted from computer simulations and by taking advantage of the two exact equations of integral equation theory. This still remains a cumbersome task, but it is much less formidable than the computation through the formal definition and it is rather feasible with modern computational resources. Part of the complexity stems from the fact that two simulation methods need to be simultaneously employed to indirectly extract bridge functions in the full range\,\cite{bridgLJ1,Yukawiso}.

The \emph{Ornstein-Zernike inversion method} is based on the fact that the radial distribution function can be directly extracted from computer simulations, being equal to the probability density of finding a particle at a distance from a reference particle relative to the probability density for an ideal gas\,\cite{theorybo}. The discretization of this definition leads to the histogram method. With knowledge of the radial distribution function, the direct correlation function can be computed from the OZ equation and the bridge function can then be computed from the closure condition,
\begin{equation}
B(r)=\ln{[g(r)]}-g(r)+\beta{u}(r)+c(r)+1\,\label{OZclosure1}\,.
\end{equation}
The OZ inversion method is rather straightforward and requires input from standard equilibrium simulations. However, the omnipresent uncertainties in the extraction of the radial distribution function propagate largely augmented to the computation of the bridge function, which necessitates very accurate simulations with a large number of particles and statistically independent configurations\,\cite{Yukawiso}. Furthermore, the OZ inversion method cannot be employed for the reliable computation of bridge functions within the entire range. The extraction of the radial distribution functions with the histogram method unavoidably leads to poor collected statistics at short distances that lie within the so-called correlation void, that can be loosely defined as $\displaystyle\mathrm{arg}_{r}\{g(r)\ll1\}$ or $\displaystyle\mathrm{arg}_{r}\{g(r)\simeq0\}$. This issue does not bear any consequences for the radial distribution function (which is practically zero) or the direct correlation function [which cannot be affected by uncertainties in the infinitesimally small $g(r)$ values], but is crucial for the bridge function courtesy of the presence of the logarithmic term in the closure condition\,\cite{Yukawiso}.

The \emph{cavity distribution method} is based on the fact that the cavity distribution function can be directly extracted from computer simulations, being equal to the radial distribution function for a tagged particle pair whose mutual interaction is suppressed that are dissolved at infinite dilution in a system where all other pair interactions remain the same\,\cite{Hansenbo,liquidbo}. The cavity distribution function remains continuous even if the interaction potential is discontinuous or diverges and acquires large but finite values near the origin $r=0$\,\cite{theorybo,Yukawiso}. In other words, by merging the radial distribution function and the interaction potential in its definition, the cavity distribution function eliminates their inherent pathologies close to the contact point. With knowledge of the direct correlation function from the OZ inversion method, the bridge function can be computed from the closure condition that reads as
\begin{equation}
B(r)\simeq\ln{[y(r)]}+c(r)+1\,.\label{OZclosure2}
\end{equation}
within the correlation void where $g(r)\simeq0$. However, the cavity distribution function increases by many orders of magnitude from the edge of the correlation void up to the origin $r=0$, which implies that uniform sample statistics are rather impossible to acquire for non-interacting tagged particles\,\cite{Yukawiso}. Hence, the major challenge in the implementation of this method is associated with the design of an interaction potential for the tagged pair that homogenizes the statistics within the entire correlation void. The artificial statistical bias that characterizes the simulated system of $N-2$ standard particles and $2$ tagged particles is known and can then be removed, so that the statistical weights that correspond to the static correlations of the physical system of $N$ standard particles are ultimately extracted. Overall, we have\,\cite{bridgLJ1,Yukawiso}
\begin{equation}
y(r)=C\exp{[\beta{\psi}(r)]}g_{\mathrm{sim}}^{12}(r)\,.\label{cavityequation}
\end{equation}
where $g_{\mathrm{sim}}^{12}(r)$ is the radial distribution function of the tagged pair that is extracted with the histogram method, $\beta\psi(r)$ is the tagged pair interaction potential and $C$ is a proportionality constant that is determined by the cavity distribution function continuity. It should be emphasized that only the tagged particles yield useful statistics, thus specially designed cavity simulations can feature a rather small number of particles but should feature a very large number of statistically independent configurations\,\cite{Yukawiso}. Note also that, due to the large cavity values within the correlation void, it is customary to work with logarithms. The definition of $y_{\mathrm{sim}}(r)=\exp{[\beta{\psi}(r)]}g_{\mathrm{sim}}^{12}(r)$ leads to
\begin{equation}
B(r)\simeq\ln{[y_{\mathrm{sim}}(r)]}+\ln{C}+c(r)+1\,.\label{OZclosure4}
\end{equation}
Finally, a simple methodology was recently proposed that streamlines the design of the externally controlled tagged pair potential so that near-uniform tagged pair statistics are acquired in the correlation void\,\cite{Yukawiso}. It is based on the decomposing the tagged potential into windowing and biasing components. The windowing part constrains the tagged pair in overlapping sub-intervals of the correlation void without affecting dynamics within each confinement range and can be realized by trial-and-error, while the biasing part ensures statistical uniformity within each window and can be successively optimized with input from cavity simulations of increasing duration.

\section{The OCP bridge function extraction from MD simulations}\label{sec:extraction}

\noindent One-component plasmas (OCP) are model systems which consist of classical point particles that are immersed in a rigid charge neutralizing background and interact via the Coulomb pair potential $u(r)=(Q^2/r)$ with $Q$ the particle charge\,\cite{Gaborboo,OCPrevBH,IchiRev1}. The presence of this inert neutralizing background guarantees the existence \& uniqueness of the thermodynamic limit and the stability of the classical OCP\,\cite{OCPrevBH}. The thermodynamic states of the classical OCP are fully specified by a single dimensionless parameter, since the non-ideal Helmholtz free energy depends on a specific density ($n$) and temperature ($T$) combination\,\cite{Helmholz}: the coupling parameter $\Gamma=\beta{Q}^2/d$ where $d=[4\pi{n}/3]^{-1/3}$ is the Wigner-Seitz radius and $\beta=1/(k_{\mathrm{B}}{T})$ the thermodynamic beta\,\cite{OCPrevBH,IchiRev1}. The OCP undergoes a liquid-to-bcc phase transition at $\Gamma_{\mathrm{m}}\simeq171.8$\,\cite{sizecor4} and its structural correlations undergo a crossover from monotonic decay to exponentially damped oscillatory decay at $\Gamma_{\mathrm{K}}\simeq1.12$\,\cite{Kirkwood}. Analysis of the poles of the Fourier transformed total correlation function has revealed that at the crossover point two imaginary poles coalesce to generate a conjugate pair of complex poles, \emph{i.e.} $\Gamma_{\mathrm{K}}$ is the so-called Kirkwood point\,\cite{Kirkwood}. For $\Gamma<\Gamma_{\mathrm{K}}$, the hypernetted chain (HNC) approach which drops the bridge function contribution is known to yield excellent results\,\cite{HNCgood1}.

In this section, we shall indirectly extract the OCP bridge functions of $17$ thermodynamic states that are uniformly distributed between the crystallization point and the Kirkwood point. To be more specific, the OCP states of interest are $\Gamma=10,\,20,\,30,\,...\,160,\,170$. This investigation covers the entire stable OCP phase diagram range of interest, since the effect of the bridge functions is greatly diminished below the Kirkwood point. In the following, reduced units $x=r/d$ and $q=kd$ will be mainly employed for the real \& reciprocal space, which lead to the interaction potential $\beta{u}(x)=\Gamma/x$ or $\beta\widetilde{u}(q)=4\pi\Gamma/q^2$.

\subsection{Intermediate and long range extraction}\label{subsec:extraction_long}

\noindent  The standard canonical (NVT) MD simulations are carried out with the LAMMPS software\,\cite{LAMMPSre}. The long-range nature of Coulomb interactions is handled with the Ewald decomposition\,\cite{PPPMref1} that is implemented with the particle-particle particle-mesh (PPPM) technique\,\cite{PPPMref2,PPPMref3}. Simulations feature $2^{20}$ time-steps for equilibration, $2^{23}$ time-steps for statistics and $2^7$ time-step saving period which ensures that statistically independent configurations are only postprocessed\,\cite{independ}. This leads to $M=2^{16}(=65536)$ for the number of uncorrelated configurations. The simulated particle number is $N=54872$ yielding $L/d=60$ for the length of the primitive cell of the periodic cubic box. Note that $N,M$ are chosen so that the OZ inversion method leads to accurate bridge functions for $r/d\geq1.25$.

The radial distribution functions are subject to statistical errors owing to the finite simulation duration, grid errors due to the finite histogram bin width, size errors due to the finite particle number and tail errors due to the finite simulation box length\,\cite{errorsHS,Yukawiso}. These uncertainties propagate augmented from the radial distribution function to the bridge function in the course of the OZ inversion method. Similar to Ref.\cite{Yukawiso}, statistical errors will be quantified, grid errors will be minimized and size errors will be corrected, while tail errors are negligible.

\emph{Statistical errors} are quantified with the application of a block averaging procedure which divides the dataset of $M$ uncorrelated configurations into $N_{\mathrm{b}}$ blocks each containing $N_{\mathrm{g}}$ configurations. In order to ensure acquisition of sufficiently smooth block radial distribution functions that lead to the computation of meaningful block bridge functions, $N_{\mathrm{g}}$ should be large. In order to ensure availability of a large sample of block bridge functions so that statistical deviations can be accurately calculated, $N_{\mathrm{b}}$ should also be large. Exhaustive empirical testing of various $N_{\mathrm{b}},N_{\mathrm{g}}$ combinations confirmed the earlier conclusion that $N_{\mathrm{b}}=N_{\mathrm{g}}=256$ is the near-optimal choice\,\cite{Yukawiso}.

\begin{figure*}
	\centering
	\includegraphics[width=6.15in]{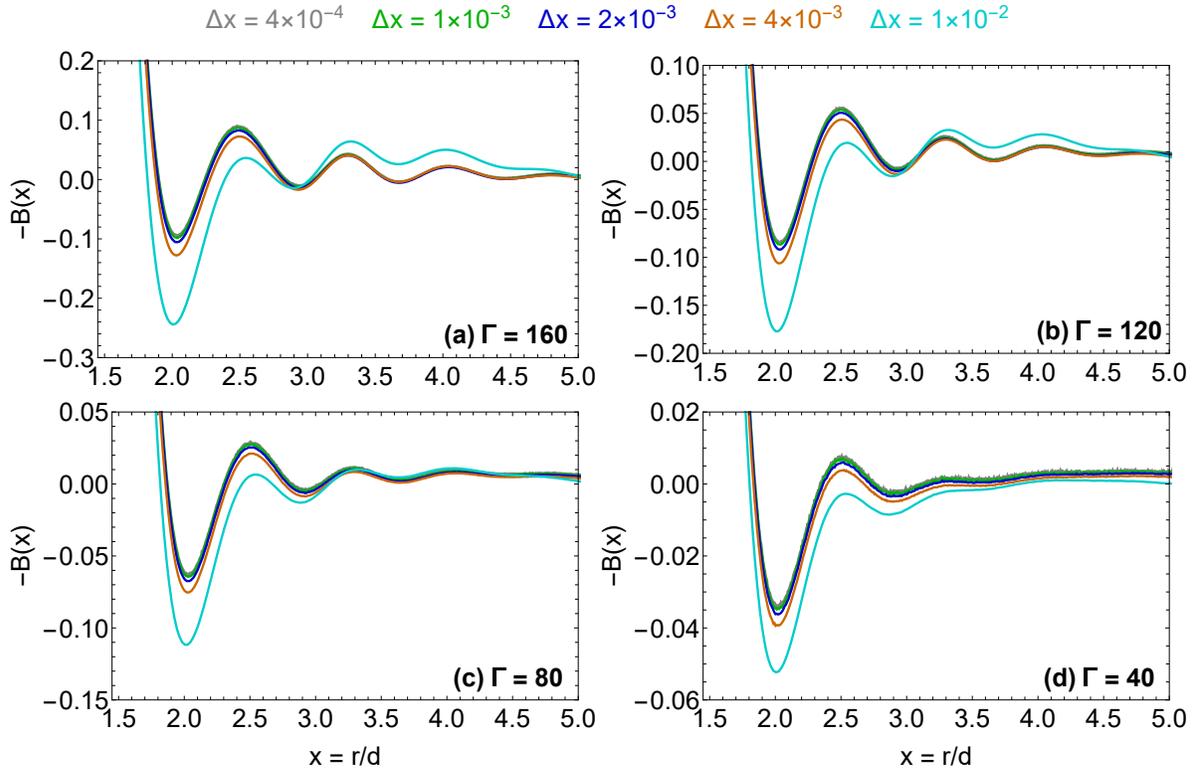}
	\caption{\emph{Grid errors} in the computation of OCP bridge functions at the state points (a) $\Gamma=160$, (b) $\Gamma=120$, (c) $\Gamma=80$ and (d) $\Gamma=40$ with input from accurate standard MD simulations. Determination for five different histogram bin widths, namely $\Delta{r}/d=0.0004,\,0.001,\,0.002,\,0.004,\,0.01$. For the probed bin widths, regardless of the coupling parameter, the global extremum magnitude increases as the bin width increases. However, it is evident that a near-optimal bin width emerges below which the OCP bridge functions overlap. Unless the chosen bin widths are large, \emph{i.e.} $\Delta{r}/d\gtrsim0.01$, the grid errors are rather small. Note also that, as the coupling parameter increases, the bridge function dependence on the bin width becomes noticeably stronger.}\label{fig:grid_errors}
\end{figure*}

\emph{Grid errors} can be made negligible compared to the statistical errors by controlling the bin width of the histograms that are employed for the extraction of the radial distribution function. The exact value is determined by an empirical analysis of the dependence of the bridge function on the bin width\,\cite{errorsHS}. It has been shown that, as the bin width decreases, the bridge function becomes independent of its value\,\cite{Yukawiso}. The near-optimal value corresponds to the largest bin width for which convergence is observed, since further bin width reduction would not affect the average bridge function but would increase the average bridge function fluctuations due to the statistical error increase\,\cite{Yukawiso}. For all the OCP state points of interest, the near-optimal bin width is close to $\Delta{r}/d=0.002$. Fig.\ref{fig:grid_errors} features characteristic examples for $4$ OCP states.

\begin{figure}
	\centering
	\includegraphics[width=3.40in]{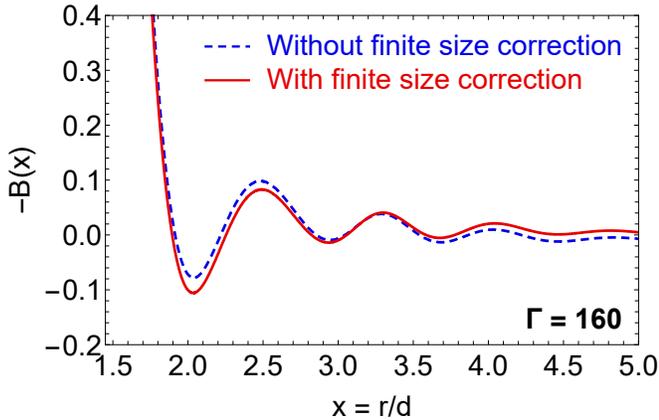}
	\caption{\emph{Finite size errors} in the computation of OCP bridge function at the state point $\Gamma=160$ with input from accurate standard MD simulations. Results with and without the simplified Lebowitz-Percus correction. The correction slightly affects the oscillatory decay region and mainly manifests itself at the asymptotic limit which now properly converges to zero.}\label{fig:finite_size_errors}
\end{figure}

\emph{Finite size errors} are compensated for by applying the Lebowitz-Percus expression, which corrects (up to second order) for the effect of particle number fluctuations that are suppressed in the canonical ensemble\,\cite{sizecor1,sizecor2}. Since all the OCP thermodynamic states of interest lie between the Kirkwood point and the crystallization point, the simplified version of the Lebowitz-Percus correction can be employed that neglects the first- and second-order coupling parameter derivative of the radial distribution function\,\cite{Yukawiso}. It simply reads as $g_{\mathrm{c}}(x)=\left[1+(\chi_{\mathrm{T}}/N)\right]g_{\mathrm{MD}}(x)$ where $g_{\mathrm{c}}(x)$ is the corrected radial distribution function, $g_{\mathrm{MD}}(x)$ the NVT-MD extracted radial distribution function, $\chi_{\mathrm{T}}$ the reduced isothermal compressibility\,\cite{Yukawiso,sizecor3}. Given the availability of an accurate OCP internal energy equation of state\,\cite{sizecor4}, the $\chi_{\mathrm{T}}$ quantity is calculated from the internal energy expression after performing the necessary thermodynamic integrations and differentiations and is not extracted from the hypervirial route\,\cite{Yukawiso}. The magnitude of the Lebowitz-Percus correction is very small, but it has an aperiodic character\,\cite{Yukawiso}. As a consequence, it has an observable effect on the extracted bridge functions, mainly at the largest coupling parameters, as deduced from Fig.\ref{fig:finite_size_errors}. More specifically, it allows to restore the correct asymptotic behavior of the bridge function.

\begin{table}
	\caption{Pad\'e approximant strategy for the enforcement of the exact long wavelength limit of the OCP structure factor. Specification of the $\sigma_1$,\,$\sigma_2$ coefficients present in the Pad\'e approximant which artificially extends the compressibility sum rule to larger wavenumbers. The $\sigma_1$,\,$\sigma_2$ values are determined by least square fitting the approximant to the MD data within the interval $[1.0,2.4]$. The bridge functions are insensitive to small perturbations of the endpoints of the selected interval. Results for the $17$ OCP state points of interest.}\label{padetable}
	\centering
	\begin{tabular}{cccccccccccc}\hline
 $\Gamma$        & $\sigma_1$           &  $\sigma_2$       & $\Gamma$       & $\sigma_1$           & $\sigma_2$          \\ \hline
 \,\,\,170\,\,\, & \,\,\,0.001243\,\,\, & \,\,\,2.281\,\,\, & \,\,\,80\,\,\, & \,\,\,0.001044\,\,\, & \,\,\,1.032\,\,\,   \\
 \,\,\,160\,\,\, & \,\,\,0.001223\,\,\, & \,\,\,2.142\,\,\, & \,\,\,70\,\,\, & \,\,\,0.001059\,\,\, & \,\,\,0.895\,\,\,   \\
 \,\,\,150\,\,\, & \,\,\,0.001248\,\,\, & \,\,\,2.004\,\,\, & \,\,\,60\,\,\, & \,\,\,0.000965\,\,\, & \,\,\,0.759\,\,\,   \\
 \,\,\,140\,\,\, & \,\,\,0.001297\,\,\, & \,\,\,1.868\,\,\, & \,\,\,50\,\,\, & \,\,\,0.000893\,\,\, & \,\,\,0.624\,\,\,   \\
 \,\,\,130\,\,\, & \,\,\,0.001233\,\,\, & \,\,\,1.726\,\,\, & \,\,\,40\,\,\, & \,\,\,0.000858\,\,\, & \,\,\,0.492\,\,\,   \\
 \,\,\,120\,\,\, & \,\,\,0.001145\,\,\, & \,\,\,1.584\,\,\, & \,\,\,30\,\,\, & \,\,\,0.000749\,\,\, & \,\,\,0.361\,\,\,   \\
 \,\,\,110\,\,\, & \,\,\,0.001152\,\,\, & \,\,\,1.446\,\,\, & \,\,\,20\,\,\, & \,\,\,0.000666\,\,\, & \,\,\,0.233\,\,\,   \\
 \,\,\,100\,\,\, & \,\,\,0.001111\,\,\, & \,\,\,1.307\,\,\, & \,\,\,10\,\,\, & \,\,\,0.000307\,\,\, & \,\,\,0.111\,\,\,   \\
 \,\,\, 90\,\,\, & \,\,\,0.001091\,\,\, & \,\,\,1.169\,\,\, & \,\,\,-\,\,\,  & \,\,\,-\,\,\,        & \,\,\,-\,\,\,       \\ \hline
	\end{tabular}
\end{table}

\begin{figure*}
	\centering
	\includegraphics[width=6.05in]{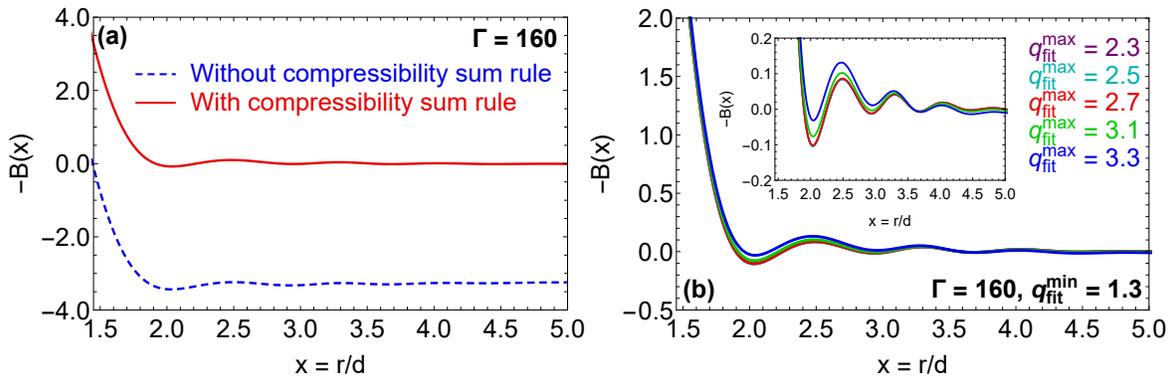}
	\caption{Importance of the \emph{exact structure factor long wavelength limit} in the extraction of OCP bridge functions. \textbf{(a)} The OCP bridge function at the state point $\Gamma=160$ with and without imposing the compressibility sum rule with the Pad\'e approximant strategy. The respective bridge functions are displaced by $\sim3.5$ in the entire reliable extraction interval. \textbf{(b)} Parametric scan of the sensitivity of the OCP bridge function at the state point $\Gamma=160$ to the upper endpoint $q_{\mathrm{fit}}^{\mathrm{max}}$ of the interval $[q_{\mathrm{fit}}^{\mathrm{min}},q_{\mathrm{fit}}^{\mathrm{max}}]$ utilized to fit the $\sigma_1$,\,$\sigma_2$ coefficients of the Pad\'e approximant strategy. The lower endpoint is kept constant $q_{\mathrm{fit}}^{\mathrm{min}}=1.3$ while the upper endpoint varies, \emph{i.e.}, $q_{\mathrm{fit}}^{\mathrm{max}}=2.3,2.5,2.7,3.1,3.3$. The (negative) bridge function is insensitive to $q_{\mathrm{fit}}^{\mathrm{max}}$ for values within $2.3-2.7$, but is displaced upwards in the intermediate range and downwards in the long range as $q_{\mathrm{fit}}^{\mathrm{max}}$ increases beyond $2.7$.}\label{fig:CSR_rule}
\end{figure*}

\begin{figure*}
	\centering
	\includegraphics[width=6.05in]{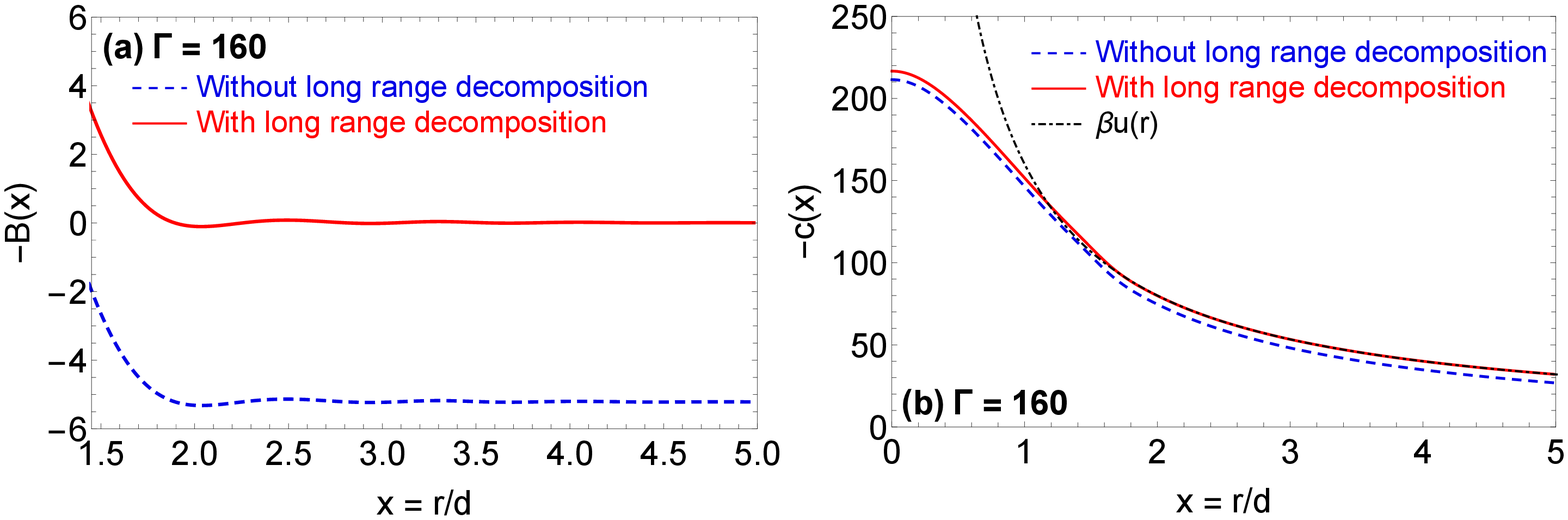}
	\caption{Importance of the \emph{long range decomposition} of the direct correlation function in the extraction of OCP bridge functions. \textbf{(a)} The OCP bridge function at the state point $\Gamma=160$ with and without employing Ng's technique of long range decomposition. The respective bridge functions are displaced by $\sim5.0$ in the entire reliable extraction interval. \textbf{(b)} The OCP direct correlation function at the state point $\Gamma=160$ with and without employing Ng's technique of long range decomposition. The respective direct correlation functions are displaced by $\sim5.0$ in the entire reliable extraction interval. This displacement directly propagates to the respective bridge functions via the non-linear closure equation. It is also evident that the long range decomposition allows the direct correlation function to obtain its exact asymptotic limit $c(x)=-\beta{u}(x)$.}\label{fig:LRD_technique}
\end{figure*}

Moreover, it is necessary to ensure that the \emph{compressibility sum rule} is exactly satisfied by the static structure factor. For the quantum and the classical OCP, the compressibility sum rule reads as $S(q\to0)=q^2/(3\Gamma+\mu_{\mathrm{T}}q^2)$ where $\mu_{\mathrm{T}}=1/\chi_{\mathrm{T}}$ is the reduced inverse isothermal compressibility\,\cite{OCPrevBH}. Two different methodologies have been devised to impose the exact long wavelength limit of the structure factor: \textbf{(i)} A Pad\'e approximant strategy, where the structure factor is assigned the form
\begin{equation*}
S(q)=
    \begin{cases}
    \displaystyle\frac{q^2+\sigma_1q^4}{3\Gamma+\mu_{\mathrm{T}}q^2+\sigma_2q^4}\,,\,\, {q}\leq{q}_{\mathrm{fit}}^{\mathrm{min}}  \\
    S_{\mathrm{MD}}(q)\,,\qquad\qquad\quad {q}\geq{q}_{\mathrm{fit}}^{\mathrm{min}}
    \end{cases}
\end{equation*}
where $S_{\mathrm{MD}}(q)$ is the structure factor that results from the extracted radial distribution function (after correcting for the finite size errors), $\sigma_1,\,\sigma_2$ are the unknown Pad\'e coefficients that are adjusted by fitting the two $S(q)$ branches within the range $q\in[{q}_{\mathrm{fit}}^{\mathrm{min}},{q}_{\mathrm{fit}}^{\mathrm{max}}]$ and ${q}_{\mathrm{fit}}^{\mathrm{min}},\,{q}_{\mathrm{fit}}^{\mathrm{max}}$ are free parameters that are determined in a trial-and-error fashion after inspecting the bridge function convergence. \textbf{(ii)} A switching function strategy, where the structure factor is assigned the form
\begin{align*}
S(q)&=L(q)S_{\mathrm{MD}}(q)+[1-L(q)]\displaystyle\frac{q^2}{3\Gamma+\mu_{\mathrm{T}}q^2}\,,\\
L(q)&=\frac{1}{2}\left\{1+ \mathrm{erf}[\alpha_1(q-\alpha_2)]\right\}\,,
\end{align*}
where $L(q)$ is a sigmoid switching function with $\mathrm{erf}(\cdot)$ the error function, $\alpha_1,\,\alpha_2$ are free parameters that are determined in a trial-and-error fashion after inspecting the bridge function convergence. It should be confirmed that the values of the free parameters for the aforementioned two strategies do not introduce any bias in the extracted bridge functions. Hence, extensive parametric scans were carried out over various $({q}_{\mathrm{fit}}^{\mathrm{min}},{q}_{\mathrm{fit}}^{\mathrm{max}})$ and $(\alpha_1,\alpha_2)$ combinations. It was deduced that the bridge functions are nearly insensitive to the $({q}_{\mathrm{fit}}^{\mathrm{min}},{q}_{\mathrm{fit}}^{\mathrm{max}})$ parameters provided that ${q}_{\mathrm{fit}}^{\mathrm{min}}\in[0.8,1.1]$ and ${q}_{\mathrm{fit}}^{\mathrm{max}}-{q}_{\mathrm{fit}}^{\mathrm{min}}\in[1.0,1.8]$, that the bridge functions are nearly insensitive to the $(\alpha_1,\alpha_2)$ parameters provided that $\alpha_1\in[6.0,10]$ and $\alpha_2\in[0.6,0.8]$, that for both strategies bridge functions become less sensitive to the free parameters as the coupling strength decreases and that both strategies result to identical bridge functions for the optimal free parameters. In what follows, the Pad\'e approximant strategy has been preferred for enforcement of the exact structure factor long wavelength limit with ${q}_{\mathrm{fit}}^{\mathrm{min}}=1.0$ and ${q}_{\mathrm{fit}}^{\mathrm{max}}=2.4$. For all state points of interest, the fitted $\sigma_1,\,\sigma_2$ Pad\'e coefficients can be found in Table \ref{padetable}. See Fig.\ref{fig:CSR_rule}a for an illustration of the importance of the compressibility sum rule in the extraction of bridge functions and Fig.\ref{fig:CSR_rule}b for the bridge function dependence on  $({q}_{\mathrm{fit}}^{\mathrm{min}},{q}_{\mathrm{fit}}^{\mathrm{max}})$ during a parametric scan.

Furthermore, due to the long-range nature of Coulomb interactions, respectable numerical errors are introduced during the inverse Fourier transform that intervenes in the computation of the real-space direct correlation function $c(x)$. These numerical errors have a strong impact on the bridge function and can be avoided by utilizing the \emph{long range decomposition} technique proposed by Ng\,\cite{LRdecomp}. In particular, a $c(x)=c_{\mathrm{s}}(x)+c_{\mathrm{l}}(x)$ decomposition of the direct correlation function is assumed, where $c_{\mathrm{s}}(x)$ is the short range part and $c_{\mathrm{l}}(x)=-\beta{u}_{\mathrm{l}}(x)$ is the long range part in view of the exact asymptotic condition $c(x)=-\beta{u}(x)$\,\cite{Hansenbo}. The function $-\beta{u}_{\mathrm{l}}(x)=-(\Gamma/x)\mathrm{erf}(\alpha{x})$ with $\alpha=1.08$ well approaches the direct correlation function in an extended range and possesses an analytical Fourier transform $\beta\widetilde{u}_{\mathrm{l}}(q)=(4\pi\Gamma/q^2)\exp{\left[-q^2/(4\alpha^2)\right]}$. Hence, only the inverse Fourier transform of the finite short-ranged small-magnitude $c_{\mathrm{s}}(x)$ function needs to be computed numerically, which leads to a drastic reduction of the associated systematic errors\,\cite{LRdecomp}. See Fig.\ref{fig:LRD_technique} for an illustration of the importance of the long range decomposition in the determination of the OCP bridge function.

\begin{figure*}
	\centering
	\includegraphics[width=6.35in]{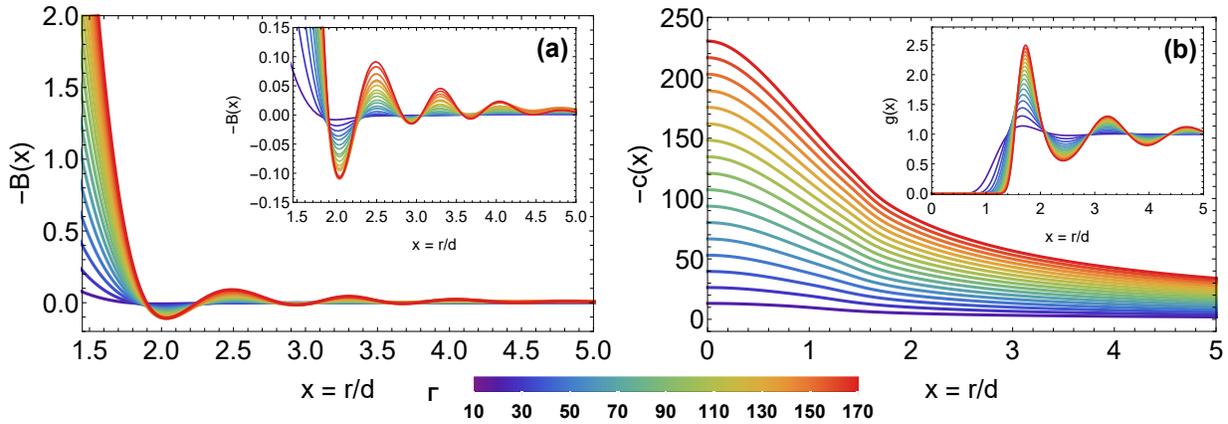}
	\caption{(a) (Main and zoomed-in inset) Color plots that feature the OCP bridge functions in the intermediate and long range $1.5\leq{x}\leq5$ for all $17$ state points of interest, as computed from the OZ inversion method with input from accurate standard MD simulations. (b) Color plots that feature the OCP direct correlation functions (main) and OCP radial distribution functions (inset) in the entire range $0\leq{x}\leq5$ for all $17$ state points of interest, as computed from the OZ inversion method with input from accurate standard MD simulations.}\label{fig:bridge_long_color}
\end{figure*}

Taking the above OCP peculiarities and the uncertainty analysis into consideration, the computation of the OCP bridge functions with the OZ inversion method and input from standard NVT MD simulations proceeds in the following steps: \textbf{(i)} The radial distribution function is
extracted from MD simulations using the histogram method with a small constant $\Delta{r}/d=0.002$ bin width so that grid errors are minimized. \textbf{(ii)} The reduced isothermal compressibility is computed from the internal energy equation of state and the Lebowitz-Percus correction is applied in order to dispose of finite size errors. \textbf{(iii)} Fast Fourier Transforms (FFT) are employed to compute the static structure factor $S(q)=1+n\widetilde{H}(q)$. \textbf{(iv)} The Pad\'e approximant strategy with ${q}_{\mathrm{fit}}^{\mathrm{min}}=1.0$ and ${q}_{\mathrm{fit}}^{\mathrm{max}}=2.4$ is utilized to ensure that the compressibility sum rule is satisfied exactly. \textbf{(v)} The Fourier transform of the direct correlation function $\widetilde{C}(q)$ is computed from the Fourier transformed OZ equation, $\widetilde{C}(q)=[S(q)-1]/[nS(q)]$, and the direct correlation function $c(x)$ is computed by applying the inverse FFT together with the long range decomposition technique. \textbf{(vi)} The average bridge function is computed from the exact non-linear closure equation that can be explicitly solved in terms of the bridge function, \emph{i.e.}, $B(x)=\ln[g(x)]-g(x)+c(x)+\beta{u}(x)+1$. \textbf{(vii)} For the quantification of the statistical errors, the above steps are also applied to all block radial distribution functions $\langle{g}_i(x)\rangle_{N_{\mathrm{g}}}$ that lead to the block bridge functions $B_i(x)=B[\langle{g}_i(x)\rangle_{N_{\mathrm{g}}}]$. The $N_{\mathrm{b}}$ samples of $B_i(x)$ are employed in the well-known statistical formula for the standard deviation of the mean of the bridge function $\sigma[B(x)]$
\begin{equation*}
\sigma[B(x)]=\sqrt{\frac{1}{N_{\mathrm{b}}(N_{\mathrm{b}}-1)}\sum_{i=1}^{N_{\mathrm{b}}}\left\{B_{i}(x)-B[\langle{g}(x)\rangle_{M}]\right\}^2}\,,
\end{equation*}
where we have $B(x)=B[\langle{g}(x)\rangle_{M}]$ for the average bridge function that is computed from the average radial distribution $g(x)=\langle{g}(x)\rangle_{M}$ prior to the block separation. The statistical error is quantified, at each point, with the near-optimal combination $N_{\mathrm{b}}=N_{\mathrm{g}}=256$. The selected error bars for the statistical uncertainties correspond to $95\%$ confidence intervals.

\begin{figure*}
	\centering
	\includegraphics[width=6.80in]{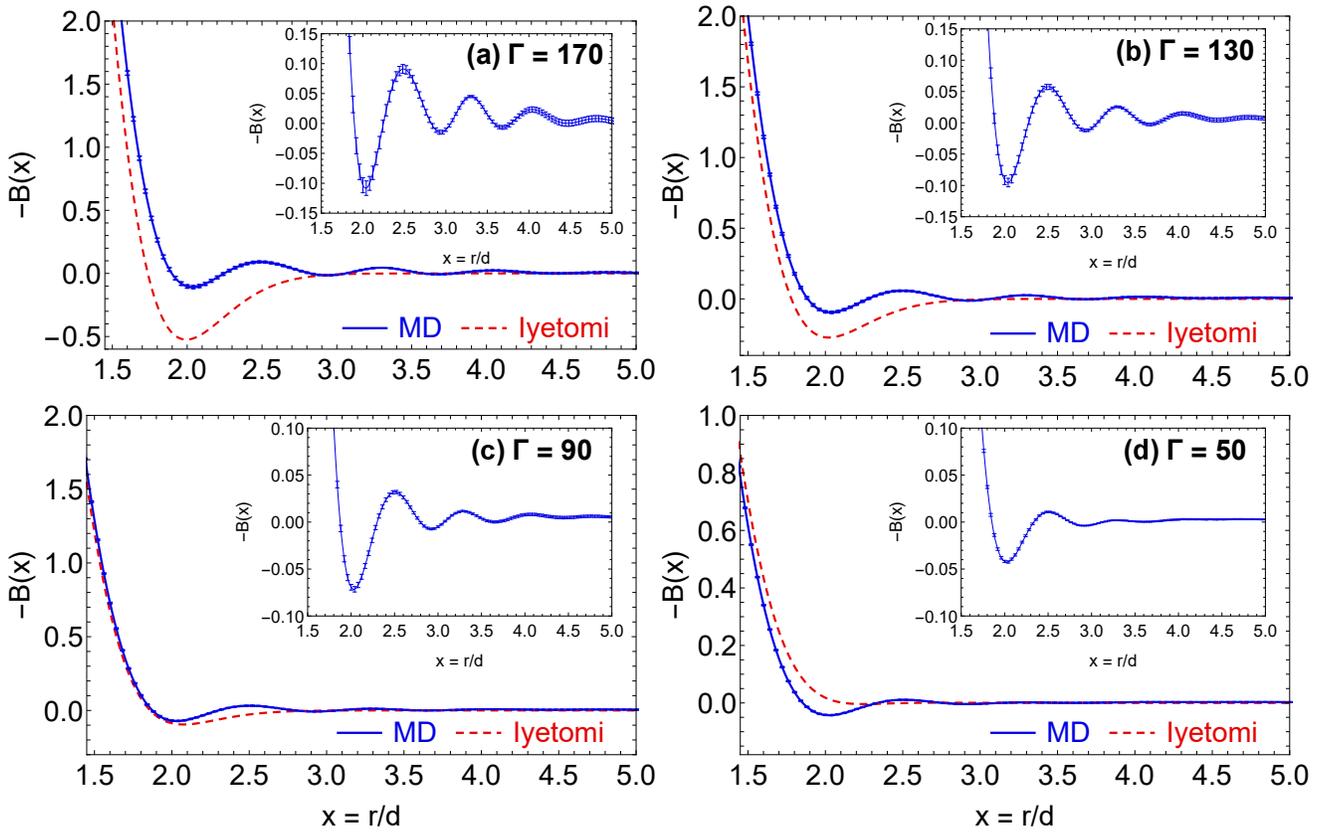}
	\caption{The OCP bridge functions in the intermediate and long range $1.5\leq{x}\leq5$ for the state points (a) $\Gamma=170$, (b) $\Gamma=130$, (c) $\Gamma=90$, (d) $\Gamma=50$. (Main) The OCP bridge functions as computed from the OZ inversion method with input from our accurate standard MD simulations versus the OCP bridge functions as calculated from the Iyetomi \emph{et al.}\,\cite{bridOCP1} parametrization. (Zoomed-in inset) The OCP bridge functions as computed from the OZ inversion method with input from our accurate standard MD simulations together with error bars that stem from the statistical uncertainties and correspond to $95\%$ confidence intervals.}\label{fig:error_bars_Iyetomi}
\end{figure*}

The extracted bridge functions $B(x)$ within the range $1.5\leq{x}\leq5.0$, for all the $17$ OCP state points of interest, are illustrated in Fig.\ref{fig:bridge_long_color}a. The OZ inversion method yields very reliable results for $x\geq1.25$, but the narrow interval $1.25\leq{x}\leq1.50$ is not included, since the necessary re-scaling due to large local magnitude of the bridge function would render the oscillatory pattern invisible. The inset features a magnification of the oscillatory decay behavior that characterizes the bridge function for $x\gtrsim1.8$. From the inset, regardless of the coupling parameter, it is evident that the OCP bridge function exhibits a number of alternating local minima and local maxima. It also turns positive at multiple periodic intervals of a roughly $0.3d$ extent, until it becomes effectively zero. This bridge function behavior is certainly a reflection of the radial distribution function behavior. Therefore, it is expected to emerge for all coupling parameters beyond the Kirkwood point, even including the metastable (supercooled) liquid states. On the other hand, an exponentially decaying bridge function behavior should be expected for coupling parameters below the Kirkwood point, although, at this point, this remains a conjecture. A review of the literature suggests that this periodic sign switching constitutes an omnipresent bridge function feature for dense liquids not only for the OCP but also for hard spheres\,\cite{bridgHS1,bridgHS2}, Lennard-Jones systems\,\cite{bridgLJ1,bridgLJ2}, soft spheres\,\cite{bridIPL1} and Yukawa systems\,\cite{Yukawiso}. This salient feature cannot be captured by the powerful variational modified hypernetted chain (VMHNC) approach that is based on the ansatz of bridge function quasi-universality\,\cite{invmeth1,ladVMHNC,rosVMHNC}, since it employs analytic Percus-Yevick bridge functions which are manifestly non-positive\,\cite{ourVMHNC}. It should be pointed out that the OCP bridge function has a smooth predictable pattern with respect to the coupling parameter $\Gamma$ within the intermediate monotonic range as well as the long oscillatory decaying range. As the coupling parameter increases, the absolute value of the intermediate range bridge function and the oscillation amplitude of the long range bridge function also monotonically increase. On the other hand, both the monotonic-to-oscillatory transition point and the oscillatory peak/trough positions are nearly independent of the coupling parameter. These clear trends suggest that a simple extrapolation- \& interpolation-friendly OCP bridge function parametrization should be viable.

The extracted direct correlation functions $c(x)$ within the range $0\leq{x}\leq5.0$, for all the $17$ OCP state points of interest, are illustrated in Fig.\ref{fig:bridge_long_color}b. It is worth pointing out that the OCP direct correlation function also has a smooth predictable pattern with respect to the coupling parameter within the entire range and that the family of the $c(x;\Gamma)$ curves is non-intersecting. It should also be noted that the direct correlation function acquires values very close to its exact asymptotic limit already from $x\sim1.6-1.8$ (close to the foot of the curve where the slope changes abruptly) and that the transition point is slightly displaced towards the origin as the coupling parameter decreases. This was also observed for the YOCP\,\cite{Yukawiso} and justifies the satisfactory performance of the soft mean spherical approximation (SMSA) for the OCP\,\cite{SMSAmina,SMSAminb}. It motivated us to fit the short range direct correlation function with the polynomial $c(x;\Gamma)=a(\Gamma)+b(\Gamma)x^2+c(\Gamma)x^3+d(\Gamma)x^5$, as suggested by the analytic SMSA OCP solution\,\cite{SMSAmina}. The SMSA-inspired fitting function proved to be very accurate for all $17$ OCP state points with mean absolute relative errors always below $0.1\%$. Furthermore, the coefficients $a(\Gamma),\,b(\Gamma),\,c(\Gamma),\,d(\Gamma)$ turned out to have a monotonic dependence on the coupling parameter and they could be easily fitted as functions of $\Gamma$. Other types of polynomial fits were also tested, but proved to be less accurate than the SMSA suggestion. For completeness, the extracted radial distribution functions $g(x)$ within the range $0\leq{x}\leq5.0$, for all the $17$ OCP state points of interest, are illustrated in the inset of Fig.\ref{fig:bridge_long_color}b. The well-known monotonic increases of the correlation void extent, of the alternating peak \& trough magnitudes and of the first coordination cell sharpness with increasing coupling parameter can be easily observed.

Finally, the extracted bridge functions $B(x)$ within the range $1.5\leq{x}\leq5.0$ are compared to the OCP bridge function parametrization of Iyetomi \emph{et al.}\,\cite{bridOCP1} (see details in the appendix) in Fig.\ref{fig:error_bars_Iyetomi}. It can be observed that: \textbf{(i)} the largest relative deviations occur in the region of the first bridge function maximum, whose magnitude and width are grossly overestimated by the Iyetomi parametrization; \textbf{(ii)} rather respectable relative deviations occur in the range of secondary maxima and minima, which are omitted in the Iyetomi parametrization; \textbf{(iii)} small relative deviations are also present in the monotonic bridge function range; \textbf{(iv)} the relative deviations decrease in a systematic manner as the coupling parameter decreases. The observed deviations in the monotonic and first maximum region as well as the decrease of deviations with decreasing coupling parameters can be explained by the large grid errors that accompany the Iyetomi \emph{et al.} bridge function extraction which stem from the use of large $\Delta{r}/d=0.04$ histogram bin widths\,\cite{bridOCP1}, compare also with Fig.\ref{fig:grid_errors}. The deviations in the secondary extrema region stem from the fact that the Iyetomi parametrization is constructed by a high-order polynomial that is multiplied by an exponential decaying function. The resulting curve features a single maximum that is followed by a prompt decay to zero. In other words, Iyetomi \emph{et al.} purposely avoided to include the description of the secondary maxima in order to avoid overcomplicated expressions. The statistical errors have been quantified in the inset of Fig.\ref{fig:error_bars_Iyetomi}. They are rather negligible, much smaller than the deviations from the Iyetomi parametrization and rather uniformly spread in the depicted region. The error bars also monotonically decrease as the coupling decreases. A systematic comparison with the Iyetomi parametrization and a comprehensive presentation of the overall uncertainties can be found in the supplementary material\,\cite{supplem1}.

\subsection{Short range extraction}\label{subsec:extraction_short}

\noindent  The specially-designed canonical (NVT) MD simulations are also carried out with the LAMMPS software\,\cite{LAMMPSre} and the Ewald sum is implemented with the PPPM method. The simulated particle number is $N=1000$, consisting of 998 standard particles and 2 tagged particles, which leads to a $L/d=16$ cubic simulation box length. The tagged particle correlation statistics are collected with histogram bins of $\Delta{r}/d=0.002$ width. The short range is split into four overlapping windows - $I_1=[0.0,0.4]$, $I_2=[0.2,0.6]$, $I_3=[0.4,1.0]$, $I_4=[0.8,1.4]$ (in $r/d$ units)- and the intermediate \& long range essentially define a fifth overlapping window - $I_5=[1.25,\infty)$. \emph{Short cavity simulations} feature $2^{19}$ time-steps for equilibration, a $2^7$ time-step saving period and $2^{19}$ time-steps for statistics regardless of the coupling parameters, which lead to $M_{\mathrm{s}}=2^{12}(=4096)$ for the number of the statistically independent particle configurations. \emph{Long cavity simulations} feature $2^{20}$ time-steps for equilibration, a configuration saving period of $2^7$ time-steps, $2^{31}$ time-steps for statistics when $\Gamma>30$ and $2^{32}$ time-steps for statistics when $\Gamma\leq30$, which yield the following numbers of statistically independent particle configurations; $M_{\mathrm{l}}=2^{24}(=16,777,216)$ when $\Gamma>30$ and $M_{\mathrm{l}}=2^{25}(=33,554,432)$ when $\Gamma=30,\,20,\,10$. Finally, the windowing component of the tagged pair interaction potential for the $n$-th overlapping interval $I_n=[b_n,c_n]$ is $\beta\chi_n(x)=100\left\{\mathrm{erf}\left[20(a_1-x)\right]+\mathrm{erf}\left[20(x-a_2)\right]\right\}\,$, where $a_1=b_n-0.1$ when $n>1$, $a_1=-0.5$ when $n=1$ and $a_2=c_n+0.1,\,\,\forall{n}$.

\begin{figure}
	\centering
	\includegraphics[width=3.40in]{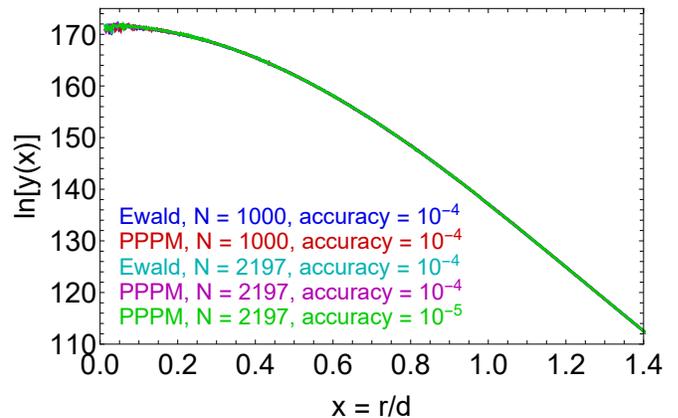}
	\caption{\emph{Finite size and Ewald sum implementation errors} in the computation of the logarithm of the OCP cavity distribution function $\ln{[y(x)]}$ within the correlation void. Results for two particle numbers ($N=1000,\,2197$), for two relative errors in the force computations ($10^{-4},10^{-5}$) and for two Ewald sum implementation techniques (the traditional Ewald method of $N^{3/2}$ scaling as well as the PPPM method of $N\log{N}$ scaling) at the $\Gamma=160$ state point. It is evident that $N=1000$ particles and a force accuracy of $10^{-4}$ in the PPPM method lead to negligible errors in the screening potential computation.}\label{fig:size_Ewald_errors}
\end{figure}

The short range bridge function $B(x)=\ln{[y_{\mathrm{sim}}(x)]}+\ln{C}+c(x)+1$ is subject to uncertainties in the screening potential extraction from the long cavity MD simulations, to uncertainties in the direct correlation function extraction from the standard MD simulations and to uncertainties in the matching constant determination\,\cite{Yukawiso}. Assuming that these uncertainty sources are statistically independent, the total bridge function uncertainties can be computed from $\sigma^2[B(x)]=\sigma^2[\ln{y_{\mathrm{sim}}(x)}]+\sigma^2[c(x)]+\sigma^2[\ln{C}]$. In a similar fashion to Ref.\cite{Yukawiso}, it has been verified that $\sigma[\ln{y_{\mathrm{sim}}(x)}]\gg\sigma[c(x)],\sigma[\ln{C}]\,\forall{x}\leq1.25$ which leads to $\sigma[B(x)]\simeq\sigma[\ln{y_{\mathrm{sim}}(x)}]\,\forall{x}\leq1.25$. It has also been verified that finite size errors and Ewald sum implementation errors are negligible, see Fig.\ref{fig:size_Ewald_errors}. Thus, the short range bridge functions are essentially only subject to statistical uncertainties in the extraction of the screening potential. The latter uncertainties are quantified by applying a block averaging procedure where the number of $N_{\mathrm{b}}$ blocks is much smaller than the number of $N_{\mathrm{g}}$ configurations per block, as a consequence of the fact that useful correlation statistics stem only from the tagged particle pair. In particular, the combination $(N_{\mathrm{b}},N_{\mathrm{g}})=(2^5,2^{19})$ when $\Gamma>30$ and the combination $(N_{\mathrm{b}},N_{\mathrm{g}})=(2^5,2^{20})$ when $\Gamma=30,\,20,\,10$, turned out to be the near-optimal choices\,\cite{Yukawiso}.

\begin{table*}
	\caption{Values of the $d_0^{n}$ and $d_1^{n}$ coefficients present in the biasing component of the tagged pair potential for the overlapping windows $I_1,\,I_2,\,I_3$ and $I_4$, as determined from the short cavity simulations and then employed in the long cavity simulations; the \enquote{0} subscript corresponds to the constant term, the \enquote{1} subscript corresponds to the linear term and the superscripts \enquote{1-4} specify the overlapping window. Results for all the $17$ OCP state points of interest. The respective values of the Ewald splitting parameter $\alpha_{\mathrm{s}}$ are also provided.}\label{biastable}
	\centering
	\begin{tabular}{cccccccccccc}\hline
$\Gamma$         & $d_0^1$             & $d_1^1$            & $d_0^2$             & $d_1^2$            & $d_0^3$             & $d_1^3$             & $d_0^4$             & $d_1^4$              & $\alpha_{\mathrm{s}}$  \\ \hline
 \,\,\,170\,\,\, & \,\,\,9.2128\,\,\, & \,\,\,-0.0877\,\,\, & \,\,\,6.1937\,\,\, & \,\,\,+3.5241\,\,\, & \,\,\,-1.3107\,\,\, & \,\,\,+8.9430\,\,\, & \,\,\,-5.3940\,\,\, & \,\,\,+8.6347\,\,\,  & \,\,\,0.4632\,\,\,     \\
 \,\,\,160\,\,\, & \,\,\,10.232\,\,\, & \,\,\,-1.3998\,\,\, & \,\,\,7.6685\,\,\, & \,\,\,+2.6468\,\,\, & \,\,\,+2.1695\,\,\, & \,\,\,+5.8998\,\,\, & \,\,\,-2.4830\,\,\, & \,\,\,+7.3717\,\,\,  & \,\,\,0.4632\,\,\,     \\
 \,\,\,150\,\,\, & \,\,\,9.5835\,\,\, & \,\,\,-1.5325\,\,\, & \,\,\,7.1782\,\,\, & \,\,\,+1.2989\,\,\, & \,\,\,+0.0302\,\,\, & \,\,\,+7.5069\,\,\, & \,\,\,-4.0295\,\,\, & \,\,\,+7.6492\,\,\,  & \,\,\,0.4650\,\,\,     \\
 \,\,\,140\,\,\, & \,\,\,9.2107\,\,\, & \,\,\,-0.2462\,\,\, & \,\,\,6.7411\,\,\, & \,\,\,+2.5474\,\,\, & \,\,\,+0.3386\,\,\, & \,\,\,+6.9836\,\,\, & \,\,\,-4.4816\,\,\, & \,\,\,+8.1610\,\,\,  & \,\,\,0.4568\,\,\,     \\
 \,\,\,130\,\,\, & \,\,\,9.6468\,\,\, & \,\,\,-1.9939\,\,\, & \,\,\,6.7408\,\,\, & \,\,\,+2.6068\,\,\, & \,\,\,+0.9334\,\,\, & \,\,\,+6.3139\,\,\, & \,\,\,-3.5504\,\,\, & \,\,\,+7.4433\,\,\,  & \,\,\,0.4569\,\,\,     \\
 \,\,\,120\,\,\, & \,\,\,9.4212\,\,\, & \,\,\,-1.0091\,\,\, & \,\,\,7.2295\,\,\, & \,\,\,+1.5578\,\,\, & \,\,\,+1.3830\,\,\, & \,\,\,+5.9210\,\,\, & \,\,\,-3.6982\,\,\, & \,\,\,+7.5858\,\,\,  & \,\,\,0.4575\,\,\,     \\
 \,\,\,110\,\,\, & \,\,\,9.5927\,\,\, & \,\,\,-1.6088\,\,\, & \,\,\,7.4653\,\,\, & \,\,\,+1.2137\,\,\, & \,\,\,+2.7367\,\,\, & \,\,\,+4.4134\,\,\, & \,\,\,-3.1008\,\,\, & \,\,\,+7.1095\,\,\,  & \,\,\,0.4559\,\,\,     \\
 \,\,\,100\,\,\, & \,\,\,9.4448\,\,\, & \,\,\,-1.0458\,\,\, & \,\,\,7.0443\,\,\, & \,\,\,+2.4022\,\,\, & \,\,\,+1.8630\,\,\, & \,\,\,+5.6496\,\,\, & \,\,\,-1.7073\,\,\, & \,\,\,+5.9338\,\,\,  & \,\,\,0.4577\,\,\,     \\
 \,\,\, 90\,\,\, & \,\,\,9.4561\,\,\, & \,\,\,-0.9844\,\,\, & \,\,\,7.7185\,\,\, & \,\,\,+0.6584\,\,\, & \,\,\,+2.6952\,\,\, & \,\,\,+4.7028\,\,\, & \,\,\,-0.6456\,\,\, & \,\,\,+5.0851\,\,\,  & \,\,\,0.4570\,\,\,     \\
 \,\,\, 80\,\,\, & \,\,\,9.6144\,\,\, & \,\,\,-1.4034\,\,\, & \,\,\,8.3187\,\,\, & \,\,\,-0.5255\,\,\, & \,\,\,+4.0379\,\,\, & \,\,\,+2.9926\,\,\, & \,\,\,+0.5337\,\,\, & \,\,\,+4.2342\,\,\,  & \,\,\,0.4475\,\,\,     \\
 \,\,\, 70\,\,\, & \,\,\,9.6208\,\,\, & \,\,\,-0.7962\,\,\, & \,\,\,7.9356\,\,\, & \,\,\,+0.2386\,\,\, & \,\,\,+3.6957\,\,\, & \,\,\,+3.6274\,\,\, & \,\,\,+0.0548\,\,\, & \,\,\,+4.7159\,\,\,  & \,\,\,0.4465\,\,\,     \\
 \,\,\, 60\,\,\, & \,\,\,9.6515\,\,\, & \,\,\,-1.1721\,\,\, & \,\,\,8.2530\,\,\, & \,\,\,-0.2686\,\,\, & \,\,\,+4.6930\,\,\, & \,\,\,+2.4704\,\,\, & \,\,\,+1.8648\,\,\, & \,\,\,+3.2985\,\,\,  & \,\,\,0.4464\,\,\,     \\
 \,\,\, 50\,\,\, & \,\,\,9.2518\,\,\, & \,\,\,+0.0579\,\,\, & \,\,\,6.9076\,\,\, & \,\,\,+2.6810\,\,\, & \,\,\,+1.9595\,\,\, & \,\,\,+5.5792\,\,\, & \,\,\,-1.0721\,\,\, & \,\,\,+5.6315\,\,\,  & \,\,\,0.4401\,\,\,     \\
 \,\,\, 40\,\,\, & \,\,\,9.8287\,\,\, & \,\,\,-2.0023\,\,\, & \,\,\,8.4183\,\,\, & \,\,\,-0.4184\,\,\, & \,\,\,+5.3728\,\,\, & \,\,\,+1.6568\,\,\, & \,\,\,+3.0018\,\,\, & \,\,\,+2.4565\,\,\,  & \,\,\,0.4409\,\,\,     \\
 \,\,\, 30\,\,\, & \,\,\,9.7423\,\,\, & \,\,\,-1.5448\,\,\, & \,\,\,8.5569\,\,\, & \,\,\,-0.6829\,\,\, & \,\,\,+6.4510\,\,\, & \,\,\,+0.2333\,\,\, & \,\,\,+4.3324\,\,\, & \,\,\,+1.3006\,\,\,  & \,\,\,0.4352\,\,\,     \\
 \,\,\, 20\,\,\, & \,\,\,0.0000\,\,\, & \,\,\,0.0000\,\,\,  & \,\,\,8.7061\,\,\, & \,\,\,-1.0834\,\,\, & \,\,\,+7.0266\,\,\, & \,\,\,-0.4753\,\,\, & \,\,\,+5.3838\,\,\, & \,\,\,+0.4122\,\,\,  & \,\,\,0.4368\,\,\,     \\
 \,\,\, 10\,\,\, & \,\,\,0.0000\,\,\, & \,\,\,0.0000\,\,\,  & \,\,\,8.9069\,\,\, & \,\,\,-1.5666\,\,\, & \,\,\,+6.7065\,\,\, & \,\,\,+0.1510\,\,\, & \,\,\,+6.2936\,\,\, & \,\,\,-0.3329\,\,\,  & \,\,\,0.4171\,\,\,     \\ \hline
	\end{tabular}
\end{table*}

\begin{table*}
	\caption{Results of the long cavity simulation matching procedure for all the $17$\,OCP state points of interest. The matching in the overlapping interval $I_5\bigcap{I}_4=[1.25,1.4]$ leads to the constant $\ln{C_4}$, the matching in the overlapping interval $I_4\bigcap{I}_3=[0.8,1.0]$ leads to the constant $\ln{C_3}$, the matching in the overlapping interval $I_3\bigcap{I}_2=[0.4,0.6]$ leads to the constant $\ln{C_2}$, the matching in the overlapping interval $I_2\bigcap{I}_1=[0.2,0.4]$ leads to the constant $\ln{C_1}$. In each matching stage, the a priori knowledge of the logarithm of the cavity distribution function in the outermost interval facilitates its determination in the inner interval, starting from the intermediate and long range region $I_5$ where the screening potential $\ln{[y(x)]}$ is available from the OZ inversion method. More specifically, the proportionality constant is determined by the least square fitting of $\ln{\left[y_{I_n}(x)/y_{\mathrm{sim}}(x)\right]}=\ln{C_{n-1}}\,\forall{x}\in{I}_{n}\bigcap{I}_{n-1}$ that allows for the determination of $y_{I_{n-1}}(x)$ from $\ln{\left[y_{I_{n-1}}(x)/y_{\mathrm{sim}}(x)\right]}=\ln{C_{n-1}}\,\forall{x}\in{I}_{n-1}$. The very low mean absolute relative deviations of $e_{\ln{C_n}}\,(<0.02\%)$ confirm the theoretical expectation that the ratio $\ln{\left[y(x)/y_{\mathrm{sim}}(x)\right]}$ is constant.}\label{matchingtable}
	\centering
	\begin{tabular}{cccccccccccc}\hline
$\Gamma$         & $\ln{C_4}$          & $e_{\ln{C_4}}$     & $\ln{C_3}$          & $e_{\ln{C_3}}$     & $\ln{C_2}$          & $e_{\ln{C_2}}$     & $\ln{C_1}$          & $e_{\ln{C_1}}$          \\ \hline
 \,\,\,170\,\,\, & \,\,\,111.268\,\,\, & \,\,\,0.01\%\,\,\, & \,\,\,104.425\,\,\, & \,\,\,0.02\%\,\,\, & \,\,\, 97.561\,\,\, & \,\,\,0.02\%\,\,\, & \,\,\,94.037\,\,\,  & \,\,\,0.01\%\,\,\,      \\
 \,\,\,160\,\,\, & \,\,\,114.639\,\,\, & \,\,\,0.01\%\,\,\, & \,\,\,108.545\,\,\, & \,\,\,0.02\%\,\,\, & \,\,\,101.716\,\,\, & \,\,\,0.02\%\,\,\, & \,\,\,98.711\,\,\,  & \,\,\,0.01\%\,\,\,      \\
 \,\,\,150\,\,\, & \,\,\,120.327\,\,\, & \,\,\,0.01\%\,\,\, & \,\,\,114.011\,\,\, & \,\,\,0.02\%\,\,\, & \,\,\,107.810\,\,\, & \,\,\,0.02\%\,\,\, & \,\,\,104.386\,\,\, & \,\,\,0.01\%\,\,\,      \\
 \,\,\,140\,\,\, & \,\,\,125.815\,\,\, & \,\,\,0.01\%\,\,\, & \,\,\,120.263\,\,\, & \,\,\,0.01\%\,\,\, & \,\,\,114.075\,\,\, & \,\,\,0.01\%\,\,\, & \,\,\,110.898\,\,\, & \,\,\,0.01\%\,\,\,      \\
 \,\,\,130\,\,\, & \,\,\,130.423\,\,\, & \,\,\,0.01\%\,\,\, & \,\,\,125.156\,\,\, & \,\,\,0.01\%\,\,\, & \,\,\,119.176\,\,\, & \,\,\,0.01\%\,\,\, & \,\,\,116.222\,\,\, & \,\,\,0.01\%\,\,\,      \\
 \,\,\,120\,\,\, & \,\,\,135.026\,\,\, & \,\,\,0.01\%\,\,\, & \,\,\,129.905\,\,\, & \,\,\,0.01\%\,\,\, & \,\,\,124.266\,\,\, & \,\,\,0.01\%\,\,\, & \,\,\,121.248\,\,\, & \,\,\,0.01\%\,\,\,      \\
 \,\,\,110\,\,\, & \,\,\,139.836\,\,\, & \,\,\,0.01\%\,\,\, & \,\,\,134.928\,\,\, & \,\,\,0.01\%\,\,\, & \,\,\,129.538\,\,\, & \,\,\,0.01\%\,\,\, & \,\,\,126.679\,\,\, & \,\,\,0.01\%\,\,\,      \\
 \,\,\,100\,\,\, & \,\,\,144.386\,\,\, & \,\,\,0.01\%\,\,\, & \,\,\,139.554\,\,\, & \,\,\,0.01\%\,\,\, & \,\,\,134.346\,\,\, & \,\,\,0.01\%\,\,\, & \,\,\,131.660\,\,\, & \,\,\,0.01\%\,\,\,      \\
 \,\,\, 90\,\,\, & \,\,\,149.089\,\,\, & \,\,\,0.01\%\,\,\, & \,\,\,144.520\,\,\, & \,\,\,0.01\%\,\,\, & \,\,\,139.744\,\,\, & \,\,\,0.01\%\,\,\, & \,\,\,136.922\,\,\, & \,\,\,0.01\%\,\,\,      \\
 \,\,\, 80\,\,\, & \,\,\,154.257\,\,\, & \,\,\,0.00\%\,\,\, & \,\,\,150.290\,\,\, & \,\,\,0.01\%\,\,\, & \,\,\,145.708\,\,\, & \,\,\,0.01\%\,\,\, & \,\,\,142.971\,\,\, & \,\,\,0.01\%\,\,\,      \\
 \,\,\, 70\,\,\, & \,\,\,158.845\,\,\, & \,\,\,0.00\%\,\,\, & \,\,\,155.057\,\,\, & \,\,\,0.01\%\,\,\, & \,\,\,150.845\,\,\, & \,\,\,0.01\%\,\,\, & \,\,\,147.958\,\,\, & \,\,\,0.01\%\,\,\,      \\
 \,\,\, 60\,\,\, & \,\,\,163.324\,\,\, & \,\,\,0.00\%\,\,\, & \,\,\,159.871\,\,\, & \,\,\,0.01\%\,\,\, & \,\,\,155.834\,\,\, & \,\,\,0.01\%\,\,\, & \,\,\,153.167\,\,\, & \,\,\,0.01\%\,\,\,      \\
 \,\,\, 50\,\,\, & \,\,\,168.502\,\,\, & \,\,\,0.01\%\,\,\, & \,\,\,165.455\,\,\, & \,\,\,0.01\%\,\,\, & \,\,\,161.199\,\,   & \,\,\,0.01\%\,\,\, & \,\,\,158.653\,\,\, & \,\,\,0.01\%\,\,\,      \\
 \,\,\, 40\,\,\, & \,\,\,172.590\,\,\, & \,\,\,0.00\%\,\,\, & \,\,\,169.817\,\,\, & \,\,\,0.01\%\,\,\, & \,\,\,166.095\,\,\, & \,\,\,0.01\%\,\,\, & \,\,\,163.788\,\,\, & \,\,\,0.01\%\,\,\,      \\
 \,\,\, 30\,\,\, & \,\,\,177.323\,\,\, & \,\,\,0.00\%\,\,\, & \,\,\,174.878\,\,\, & \,\,\,0.01\%\,\,\, & \,\,\,171.264\,\,\, & \,\,\,0.01\%\,\,\, & \,\,\,169.066\,\,\, & \,\,\,0.01\%\,\,\,      \\
 \,\,\, 20\,\,\, & \,\,\,181.781\,\,\, & \,\,\,0.00\%\,\,\, & \,\,\,179.588\,\,\, & \,\,\,0.01\%\,\,\, & \,\,\,176.190\,\,\, & \,\,\,0.01\%\,\,\, & \,\,\,83.176\,\,\,  & \,\,\,0.01\%\,\,\,      \\
 \,\,\, 10\,\,\, & \,\,\,186.396\,\,\, & \,\,\,0.00\%\,\,\, & \,\,\,184.337\,\,\, & \,\,\,0.00\%\,\,\, & \,\,\,181.347\,\,\, & \,\,\,0.01\%\,\,\, & \,\,\,88.358\,\,\,  & \,\,\,0.01\%\,\,\,      \\ \hline
	\end{tabular}
\end{table*}

In the \emph{short cavity simulations}, that are dedicated to the determination of the biasing component of the tagged pair potential so that uniform statistics can be acquired within the entire correlation void, the initial strategy focused on the application of the methodology developed in Ref.\cite{Yukawiso}. This methodology generates a truncated Gaussian series representation of increasing complexity that is obtained in an iterative manner. Although it has led to uniform statistics within few iterations for disparate pair interaction potentials (Yukawa, Lennard-Jones, exponentially repulsive, inverse power law of variable softness and hard spheres), the methodology proved to be quite computationally demanding for the OCP. As a consequence, an alternative strategy is followed that takes advantage of the availability of an analytic parametrization of the screening potential $\beta{H}_{\mathrm{O}}(x)$, which was constructed by Ogata\,\cite{bridOCP2} on the basis of MC simulations (see appendix for details). In particular, at each windowing interval and state point; a short cavity simulation is performed with the biasing component $\beta\phi_{n}(x)=\beta{H}_{\mathrm{O}}(x)$, the radial distribution of the tagged pair particles $g_{\mathrm{sim}}^{12}$ is extracted, its logarithm is interpolated with the linear function $d_0^{n}+d_1^{n}x$ and the improved biasing component can, thus, be updated as $\beta\phi_{n}(x)=\beta{H}_{\mathrm{O}}(x)+d_0^{n}+d_1^{n}x$. This strategy requires only one short cavity simulation to generate high quality uniform statistics (in contrast to the original strategy that requires three), but it is system-specific. This is anticipated, since the original strategy assumes a non-interacting tagged pair for the initial guess while the alternative strategy utilizes a very accurate readily available first guess. The values of the $d_0^{n},\,d_1^{n}$ coefficients for the $4$ windows and for the $17$ OCP state points of interest are provided in Table \ref{biastable}. It should be noted that, in the treatment of the tagged particle pair interaction by the LAMMPS software, only the short-range real-space component of the Coulomb interaction $(\Gamma/x)\mathrm{erfc}(\alpha_{\mathrm{s}}x)$ (with $\mathrm{erfc}$ the complementary error function and $\alpha_{\mathrm{s}}$ the Ewald sum splitting parameter) can be replaced by the designed interaction $\psi(x)$. This leads to an additional pair interaction term that can be absorbed in either the connecting formula between the target and simulated system or the definition of the biasing component. Adopting the latter possibility, this ultimately yields the biasing component $\beta\phi_{n}(x)=\beta{H}_{\mathrm{O}}(x)+d_0^{n}+d_1^{n}x+(\Gamma/x)\mathrm{erf}(\alpha_{\mathrm{s}}x)$. For this reason, the employed values of the Ewald sum splitting parameter $\alpha_{\mathrm{s}}$ are featured in the last column of Table \ref{biastable}.

\begin{table*}
	\caption{\emph{1st-5th column:} Results for the Widom series representation of the screening potential $\ln{[y(x)]}=y_0+y_2x^2+y_4x^4+\mathcal{O}[x^6]$\,\cite{Widomthe} for all the $17$ OCP state points of interest. The very low mean absolute relative deviations of $\epsilon_{\ln{y}}<0.092\%$ between the MD extracted $\ln{[y(x)]}$ and the fitted truncated Widom series reveal that the first three terms suffice for an accurate short range representation. Note the monotonic dependence of the $y_0,y_2,y_4$ coefficients on the coupling parameter. \emph{6th-7th column:} Comparison of the second order Widom coefficient stemming from the fitted MD simulation result $y_2$ with the exact theoretical result of Jancovici for the OCP $y_2^{\mathrm{th}}=-\Gamma/4$\,\cite{Jancothe}. The low mean absolute relative deviations between these quantities $(<2.11\%)$ are indicative of the accuracy of the indirect bridge function extraction procedure that is adopted for the short range.}\label{Widomtable}
	\centering
	\begin{tabular}{ccccc|cc}\hline
$\Gamma$         & $y_0$                 & $y_2$                 & $y_4$               & $\epsilon_{\ln{y}}$   & $y_2^{\mathrm{th}}$ &  $\epsilon_{y_2}$                         \\ \hline
 \,\,\,170\,\,\, & \,\,\,$182.306$\,\,\, & \,\,\,$-41.611$\,\,\, & \,\,\,$4.932$\,\,\, & \,\,\,$0.030\%$\,\,\, & \,\,\,$-42.5$\,\,\, &	\,\,\,$2.092\%$\,\,\,                    \\
 \,\,\,160\,\,\, & \,\,\,$171.642$\,\,\, & \,\,\,$-39.157$\,\,\, & \,\,\,$4.636$\,\,\, & \,\,\,$0.031\%$\,\,\, & \,\,\,$-40.0$\,\,\, &	\,\,\,$2.107\%$\,\,\,                    \\
 \,\,\,150\,\,\, & \,\,\,$160.989$\,\,\, & \,\,\,$-36.722$\,\,\, & \,\,\,$4.348$\,\,\, & \,\,\,$0.032\%$\,\,\, & \,\,\,$-37.5$\,\,\, &	\,\,\,$2.075\%$\,\,\,                    \\
 \,\,\,140\,\,\, & \,\,\,$150.334$\,\,\, & \,\,\,$-34.284$\,\,\, & \,\,\,$4.056$\,\,\, & \,\,\,$0.031\%$\,\,\, & \,\,\,$-35.0$\,\,\, &	\,\,\,$2.046\%$\,\,\,                    \\
 \,\,\,130\,\,\, & \,\,\,$139.663$\,\,\, & \,\,\,$-31.825$\,\,\, & \,\,\,$3.757$\,\,\, & \,\,\,$0.031\%$\,\,\, & \,\,\,$-32.5$\,\,\, &	\,\,\,$2.077\%$\,\,\,                    \\
 \,\,\,120\,\,\, & \,\,\,$128.983$\,\,\, & \,\,\,$-29.367$\,\,\, & \,\,\,$3.461$\,\,\, & \,\,\,$0.031\%$\,\,\, & \,\,\,$-30.0$\,\,\, &	\,\,\,$2.110\%$\,\,\,                    \\
 \,\,\,110\,\,\, & \,\,\,$118.311$\,\,\, & \,\,\,$-26.932$\,\,\, & \,\,\,$3.173$\,\,\, & \,\,\,$0.032\%$\,\,\, & \,\,\,$-27.5$\,\,\, &	\,\,\,$2.065\%$\,\,\,                    \\
 \,\,\,100\,\,\, & \,\,\,$107.638$\,\,\, & \,\,\,$-24.486$\,\,\, & \,\,\,$2.880$\,\,\, & \,\,\,$0.032\%$\,\,\, & \,\,\,$-25.0$\,\,\, &	\,\,\,$2.056\%$\,\,\,                    \\
 \,\,\, 90\,\,\, & \,\,\, $96.948$\,\,\, & \,\,\,$-22.034$\,\,\, & \,\,\,$2.585$\,\,\, & \,\,\,$0.033\%$\,\,\, & \,\,\,$-22.5$\,\,\, &	\,\,\,$2.071\%$\,\,\,                    \\
 \,\,\, 80\,\,\, & \,\,\, $86.263$\,\,\, & \,\,\,$-19.610$\,\,\, & \,\,\,$2.304$\,\,\, & \,\,\,$0.035\%$\,\,\, & \,\,\,$-20.0$\,\,\, &	\,\,\,$1.950\%$\,\,\,                    \\
 \,\,\, 70\,\,\, & \,\,\, $75.552$\,\,\, & \,\,\,$-17.145$\,\,\, & \,\,\,$2.005$\,\,\, & \,\,\,$0.034\%$\,\,\, & \,\,\,$-17.5$\,\,\, &	\,\,\,$2.029\%$\,\,\,                    \\
 \,\,\, 60\,\,\, & \,\,\, $64.838$\,\,\, & \,\,\,$-14.699$\,\,\, & \,\,\,$1.714$\,\,\, & \,\,\,$0.035\%$\,\,\, & \,\,\,$-15.0$\,\,\, &	\,\,\,$2.007\%$\,\,\,                    \\
 \,\,\, 50\,\,\, & \,\,\, $54.119$\,\,\, & \,\,\,$-12.260$\,\,\, & \,\,\,$1.426$\,\,\, & \,\,\,$0.034\%$\,\,\, & \,\,\,$-12.5$\,\,\, &	\,\,\,$1.920\%$\,\,\,                    \\
 \,\,\, 40\,\,\, & \,\,\, $43.371$\,\,\, & \,\,\, $-9.811$\,\,\, & \,\,\,$1.136$\,\,\, & \,\,\,$0.039\%$\,\,\, & \,\,\,$-10.0$\,\,\, &	\,\,\,$1.890\%$\,\,\,                    \\
 \,\,\, 30\,\,\, & \,\,\, $32.596$\,\,\, & \,\,\, $-7.358$\,\,\, & \,\,\,$0.846$\,\,\, & \,\,\,$0.038\%$\,\,\, & \,\,\,$-7.5$\,\,\,  &	\,\,\,$1.893\%$\,\,\,                    \\
 \,\,\, 20\,\,\, & \,\,\, $21.780$\,\,\, & \,\,\, $-4.922$\,\,\, & \,\,\,$0.568$\,\,\, & \,\,\,$0.052\%$\,\,\, & \,\,\,$-5.0$\,\,\,  &	\,\,\,$1.560\%$\,\,\,                    \\
 \,\,\, 10\,\,\, & \,\,\, $10.881$\,\,\, & \,\,\, $-2.461$\,\,\, & \,\,\,$0.284$\,\,\, & \,\,\,$0.075\%$\,\,\, & \,\,\,$-2.5$\,\,\,  &	\,\,\,$1.560\%$\,\,\,                    \\ \hline
	\end{tabular}
\end{table*}

\begin{table*}
	\caption{Comparison of the zeroth order Widom coefficient stemming from the fitted MD simulation result $y_0$ with the results of the Salpeter ion sphere expression\,\cite{salpete2}, refined Jancovici expression\,\cite{rosepyc1}, Ichimaru expression\,\cite{ichipyc3}, Rosenfeld expression\,\cite{rosepyc2} and Ogata expression\,\cite{bridOCP2}. The respective mean absolute relative deviations between the MD result and the analytic expression outcome are also reported. The MD results agree very well with the refined Jancovici \& Rosenfeld expressions and exhibit the largest deviations from the Ichimaru \& Ogata expressions.\,As expected, their deviations from the Salpeter ion sphere expression monotonically increase as the coupling parameter decreases.}\label{pyctable}
	\centering
	\begin{tabular}{cccccccccccc}\hline
$\Gamma$     & $y_0$             & $y_0^{\mathrm{ISM}}$ & $\epsilon_{\mathrm{ISM}}$ & $y_0^{\mathrm{Jan}}$ & $\epsilon_{\mathrm{Jan}}$ & $y_0^{\mathrm{Ich}}$  & $\epsilon_{\mathrm{Ich}}$ & $y_0^{\mathrm{Ros}}$ & $\epsilon_{\mathrm{Ros}}$ & $y_0^{\mathrm{Oga}}$ & $\epsilon_{\mathrm{Oga}}$  \\ \hline
 \,\,170\,\, & \,\,$182.306$\,\, & \,\,$179.69$\,\,     & \,\,$1.43\%$\,\,          &\,\,$182.445$\,\,     & \,\,$0.08\%$\,\,          &\,\,$186.168$\,\,      & \,\,$2.12\%$\,\,          & \,\,$182.948$\,\,    &	\,\,$0.35\%$\,\,          &\,\,$184.233$\,\,     & \,\,$1.06\%$\,\,           \\
 \,\,160\,\, & \,\,$171.642$\,\, & \,\,$169.12$\,\,     & \,\,$1.47\%$\,\,          &\,\,$171.794$\,\,     & \,\,$0.09\%$\,\,          &\,\,$175.325$\,\,      & \,\,$2.15\%$\,\,          &\,\,$172.283$\,\,     &	\,\,$0.37\%$\,\,          &\,\,$173.487$\,\,     & \,\,$1.07\%$\,\,           \\
 \,\,150\,\, & \,\,$160.989$\,\, & \,\,$158.55$\,\,     & \,\,$1.51\%$\,\,          &\,\,$161.140$\,\,     & \,\,$0.09\%$\,\,          &\,\,$164.475$\,\,      & \,\,$2.17\%$\,\,          &\,\,$161.611$\,\,     &	\,\,$0.39\%$\,\,          &\,\,$162.735$\,\,     & \,\,$1.08\%$\,\,           \\
 \,\,140\,\, & \,\,$150.334$\,\, & \,\,$147.98$\,\,     & \,\,$1.57\%$\,\,          &\,\,$150.482$\,\,     & \,\,$0.10\%$\,\,          &\,\,$153.618$\,\,      & \,\,$2.18\%$\,\,          &\,\,$150.932$\,\,     &	\,\,$0.40\%$\,\,          &\,\,$151.977$\,\,     & \,\,$1.09\%$\,\,           \\
 \,\,130\,\, & \,\,$139.663$\,\, & \,\,$137.41$\,\,     & \,\,$1.61\%$\,\,          &\,\,$139.819$\,\,     & \,\,$0.11\%$\,\,          &\,\,$142.752$\,\,      & \,\,$2.21\%$\,\,          &\,\,$140.246$\,\,     &	\,\,$0.42\%$\,\,          &\,\,$141.212$\,\,     & \,\,$1.11\%$\,\,           \\
 \,\,120\,\, & \,\,$128.983$\,\, & \,\,$126.84$\,\,     & \,\,$1.66\%$\,\,          &\,\,$129.152$\,\,     & \,\,$0.13\%$\,\,          &\,\,$131.877$\,\,      & \,\,$2.24\%$\,\,          &\,\,$129.552$\,\,     &	\,\,$0.44\%$\,\,          &\,\,$130.440$\,\,     & \,\,$1.13\%$\,\,           \\
 \,\,110\,\, & \,\,$118.311$\,\, & \,\,$116.27$\,\,     & \,\,$1.73\%$\,\,          &\,\,$118.480$\,\,     & \,\,$0.14\%$\,\,          &\,\,$120.993$\,\,      & \,\,$2.27\%$\,\,          &\,\,$118.849$\,\,     &	\,\,$0.45\%$\,\,          &\,\,$119.660$\,\,     & \,\,$1.14\%$\,\,           \\
 \,\,100\,\, & \,\,$107.638$\,\, & \,\,$105.70$\,\,     & \,\,$1.80\%$\,\,          &\,\,$107.801$\,\,     & \,\,$0.15\%$\,\,          &\,\,$110.099$\,\,      & \,\,$2.29\%$\,\,          &\,\,$108.137$\,\,     &	\,\,$0.46\%$\,\,          &\,\,$108.871$\,\,     & \,\,$1.15\%$\,\,           \\
 \,\, 90\,\, & \,\, $96.948$\,\, & \,\, $95.13$\,\,     & \,\,$1.88\%$\,\,          &\,\, $97.115$\,\,     & \,\,$0.17\%$\,\,          &\,\, $99.193$\,\,      & \,\,$2.32\%$\,\,          &\,\,$97.416$\,\,      &	\,\,$0.48\%$\,\,          &\,\, $98.073$\,\,     & \,\,$1.16\%$\,\,           \\
 \,\, 80\,\, & \,\, $86.263$\,\, & \,\, $84.56$\,\,     & \,\,$1.97\%$\,\,          &\,\, $86.421$\,\,     & \,\,$0.18\%$\,\,          &\,\, $88.275$\,\,      & \,\,$2.33\%$\,\,          &\,\, $86.683$\,\,     &	\,\,$0.49\%$\,\,          &\,\, $87.265$\,\,     & \,\,$1.16\%$\,\,           \\
 \,\, 70\,\, & \,\, $75.552$\,\, & \,\, $73.99$\,\,     & \,\,$2.07\%$\,\,          &\,\, $75.717$\,\,     & \,\,$0.22\%$\,\,          &\,\, $77.342$\,\,      & \,\,$2.37\%$\,\,          &\,\, $75.937$\,\,     &	\,\,$0.51\%$\,\,          &\,\, $76.445$\,\,     & \,\,$1.18\%$\,\,           \\
 \,\, 60\,\, & \,\, $64.838$\,\, & \,\, $63.42$\,\,     & \,\,$2.19\%$\,\,          &\,\, $65.001$\,\,     & \,\,$0.25\%$\,\,          &\,\, $66.394$\,\,      & \,\,$2.40\%$\,\,          &\,\, $65.177$\,\,     &	\,\,$0.52\%$\,\,          &\,\, $65.611$\,\,     & \,\,$1.19\%$\,\,           \\
 \,\, 50\,\, & \,\, $54.119$\,\, & \,\, $52.85$\,\,     & \,\,$2.34\%$\,\,          &\,\, $54.268$\,\,     & \,\,$0.28\%$\,\,          &\,\, $55.427$\,\,      & \,\,$2.42\%$\,\,          &\,\, $54.401$\,\,     &	\,\,$0.52\%$\,\,          &\,\, $54.761$\,\,     & \,\,$1.19\%$\,\,           \\
 \,\, 40\,\, & \,\, $43.371$\,\, & \,\, $42.28$\,\,     & \,\,$2.52\%$\,\,          &\,\, $43.515$\,\,     & \,\,$0.33\%$\,\,          &\,\, $44.437$\,\,      & \,\,$2.46\%$\,\,          &\,\, $43.605$\,\,     &	\,\,$0.54\%$\,\,          &\,\, $43.893$\,\,     & \,\,$1.20\%$\,\,           \\
 \,\, 30\,\, & \,\, $32.596$\,\, & \,\, $31.71$\,\,     & \,\,$2.72\%$\,\,          &\,\, $32.733$\,\,     & \,\,$0.42\%$\,\,          &\,\, $33.420$\,\,      & \,\,$2.53\%$\,\,          &\,\, $32.784$\,\,     &	\,\,$0.58\%$\,\,          &\,\, $33.001$\,\,     & \,\,$1.24\%$\,\,           \\
 \,\, 20\,\, & \,\, $21.780$\,\, & \,\, $21.14$\,\,     & \,\,$2.94\%$\,\,          &\,\, $21.906$\,\,     & \,\,$0.58\%$\,\,          &\,\, $22.365$\,\,      & \,\,$2.69\%$\,\,          &\,\, $21.930$\,\,     &	\,\,$0.69\%$\,\,          &\,\, $22.077$\,\,     & \,\,$1.36\%$\,\,           \\
 \,\, 10\,\, & \,\, $10.881$\,\, & \,\, $10.57$\,\,     & \,\,$2.86\%$\,\,          &\,\, $10.994$\,\,     & \,\,$1.04\%$\,\,          &\,\, $11.254$\,\,      & \,\,$3.43\%$\,\,          &\,\, $11.027$\,\,     &	\,\,$1.34\%$\,\,          &\,\, $11.104$\,\,     & \,\,$2.05\%$\,\,           \\ \hline
	\end{tabular}
\end{table*}

\begin{figure*}
	\centering
	\includegraphics[width=6.70in]{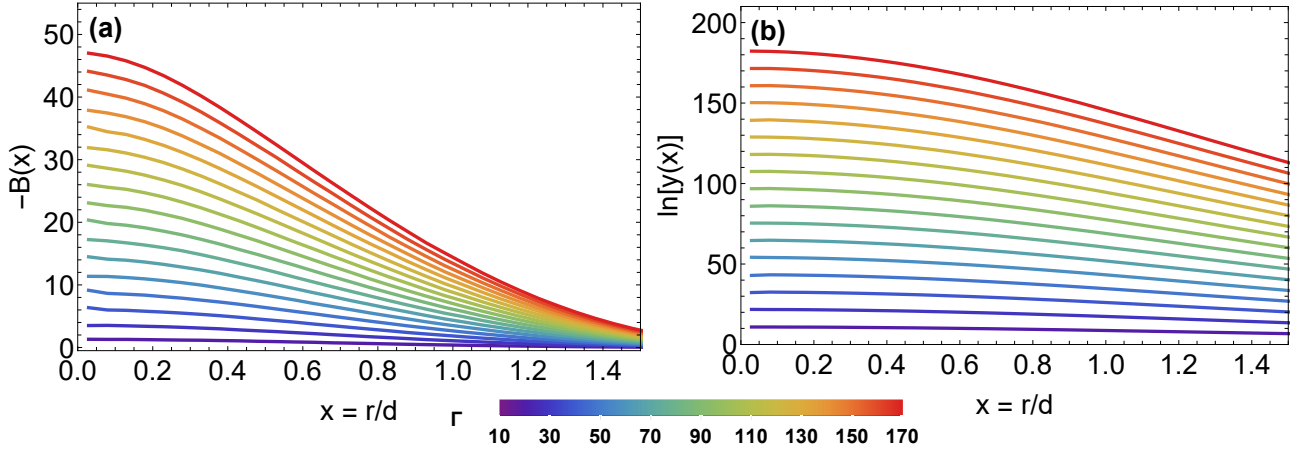}
	\caption{(a) Color plots that feature the OCP bridge functions in the short range $0\leq{x}\leq1.5$ for all $17$ state points of interest, as computed from the cavity distribution method with input from specially designed long cavity MD simulations. (b) Color plots that feature the logarithm of the OCP cavity distribution function in the short range $0\leq{x}\leq1.5$ for all $17$ state points of interest, as computed from the cavity distribution method with input from specially designed long cavity MD simulations.}\label{fig:bridge_short_color}
\end{figure*}

\begin{figure*}
	\centering
	\includegraphics[width=6.70in]{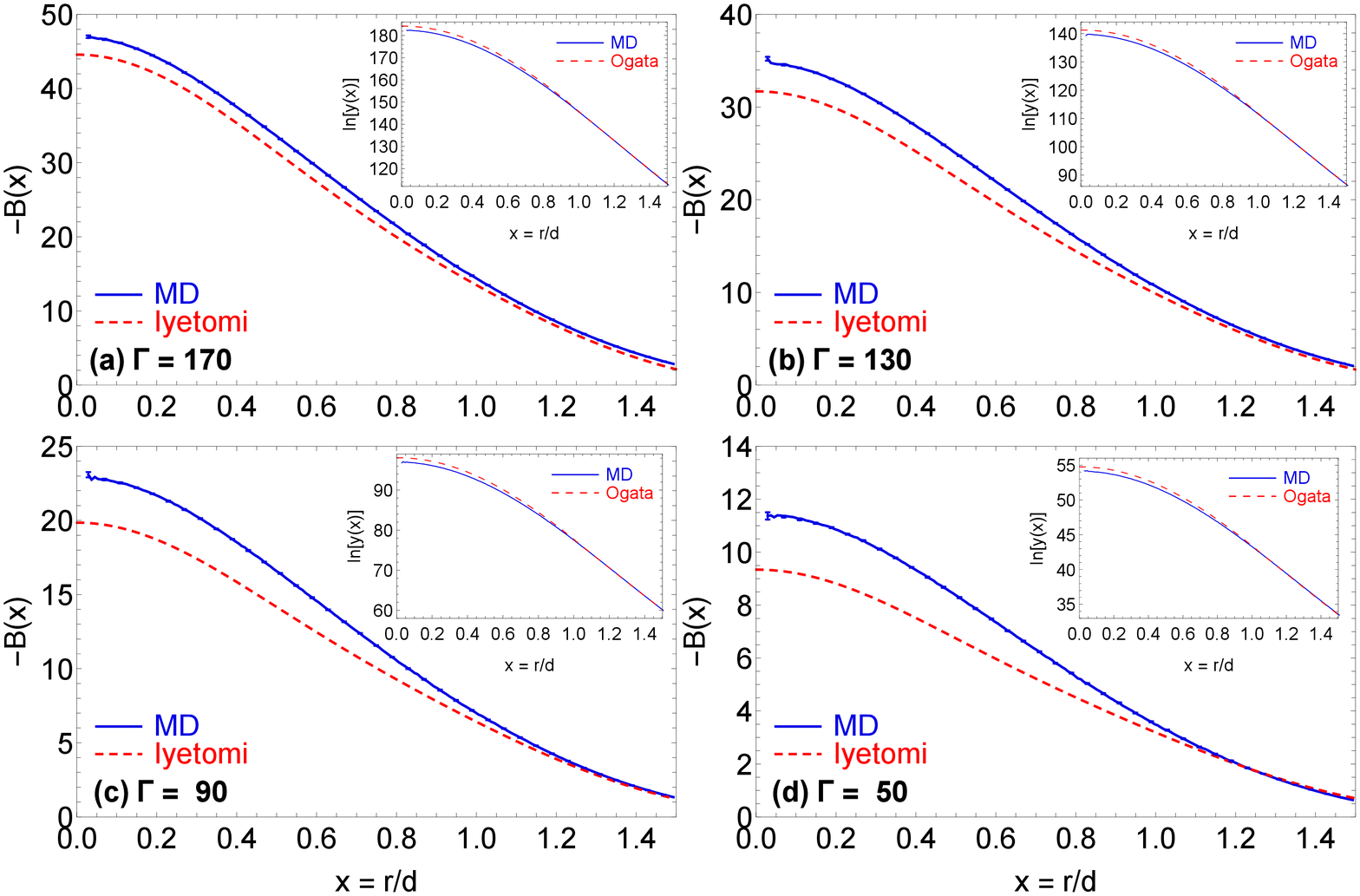}
	\caption{(Main) The OCP bridge functions in the short range $0\leq{x}\leq1.5$ for the state points (a)\,$\Gamma=170$,\,(b)\,$\Gamma=130$,\,(c)\,$\Gamma=90$, (d)\,$\Gamma=50$, as computed from the cavity distribution method with input from specially designed long cavity MD simulations versus the OCP bridge functions as calculated from the Iyetomi \emph{et al.} parametrization\,\cite{bridOCP1}. The small error bars, that originate from the statistical uncertainties, correspond to $95\%$ confidence intervals. (Inset) The OCP screening potentials in the short range $0\leq{x}\leq1.5$ for the state points (a)\,$\Gamma=170$,\,(b)\,$\Gamma=130$,\,(c)\,$\Gamma=90$,\,(d)\,$\Gamma=50$, as computed from the cavity distribution method with input from specially designed long cavity MD simulations versus the OCP screening potentials as calculated from the Ogata parametrization\,\cite{bridOCP2}.}\label{fig:error_bars_Ogata}
\end{figure*}

In the \emph{long cavity simulations}, that are dedicated to the determination of the logarithm of the cavity distribution function and thus of the short range bridge function, the aforementioned biasing component is utilized. Then, the sequential matching procedure is applied, followed in the overlapping extent of consecutive windows starting from $I_5\bigcap{I}_4$ and proceeding up to $I_2\bigcap{I}_1$, and leads to very accurate proportionality constants $\ln{C_i}$ that are provided in Table \ref{matchingtable}. \textbf{(i)} The Widom expansion for the cavity distribution function $y(x)$ reveals that it can be expressed as a $x^2$ infinite series whose coefficients alternate in sign\,\cite{Widomthe}. Therefore, it can be expected that the short range screening potential is also well approximated by an even polynomial of the general form $\ln{[y(x)]}=\sum_{i=0}^{l}y_{2i}x^{2i}$. In fact, least square fitting reveals that the first three polynomial terms $(l=2)$ suffice for an excellent agreement with the short range $\ln{[y(x)]}$ data. The coefficients $y_0,\,y_2,\,y_4$ are provided in Table \ref{Widomtable} together with the mean absolute relative errors of the fit that are extremely low for all the $17$ OCP state points of interest. \textbf{(ii)} Jancovici's theoretical result for the second order Widom OCP coefficient reads as $y_2^{\mathrm{th}}=-\Gamma/4$\,\cite{Jancothe}. The fitted MD simulation results and the exact theoretical results for the second Widom coefficient are very close, within $\sim2\%$, for all the coupling parameters, see Table \ref{Widomtable}. This serves as an independent confirmation of the accuracy of the short range bridge function extraction procedure. It is worth to emphasize that Jancovici's $y_2$ result could have been imposed on the simulation results, in the sense that $y_2=y_2^{\mathrm{th}}$ could have been assumed with the least-square fitting only concerning the zeroth and fourth order Widom OCP coefficients. Such a strategy was avoided, since it might result in uncontrolled errors in the determination of $y_0$ and $y_4$. In any case, the screening potential simulation data are preferred over the Widom series representation during the computation of the short range bridge function. \textbf{(iii)} The zeroth order Widom OCP coefficient $y_0$ has been the subject of numerous investigations, because it essentially controls the enhancement factor  of the pycnonuclear reactions in dense astrophysical plasmas due to screening effects\,\cite{ichipyc1,ichipyc2}. Importantly, the screening enhancement factor depends on $y_0$ in an exponential manner, which implies that an accurate $y_0$ determination is highly desirable\,\cite{Jancopyc}. Salpeter \& Van Horn calculated $y_0$ based on the ion sphere model and obtained values that are more appropriate in the infinite coupling parameter limit\,\cite{salpete1,salpete2}. Jancovici then computed $y_0$ by exploiting the fact that the screening potential can be expressed through the free-energy change upon fixing the positions of a particle pair in the appropriate configuration to form an interaction-site molecule, considering the limit $r\to0$ which leads to an extreme case of binary ionic mixture with $N-1$ particles of charge $Q$ and one particle of charge $2Q$, assuming the validity of the linear mixing rule for the free energy of this extreme binary ionic mixture and employing an accurate OCP equation of state\,\cite{Jancothe}. His original estimate was subsequently refined by Rosenfeld, who employed a more accurate OCP equation of state\,\cite{rosepyc1}. Ichimaru and collaborators computed $y_0$ on the basis of extra long MC simulations that were extrapolated in the short range domain within the assumption that the fourth order Widom coefficient can be approximated by zero\,\cite{bridOCP1,ichipyc3,ichipyc4}. An alternative extrapolation method was later provided by Rosenfeld that was based on the same MC data but that dropped the oversimplifying $y_4=0$ assumption\,\cite{rosepyc1,rosepyc2}. Later, Ogata directly computed $y_0$ on the basis of importance sampling MC simulations\,\cite{bridOCP2}. A comparison of the Salpeter ion sphere expression, refined Jancovici expression, Ichimaru expression, Rosenfeld expression and Ogata expression with our fitted simulation data for $y_0$ is provided in Table \ref{pyctable}. The near-excellent agreement with the independent thermodynamic expression of Jancovici further increases confidence to the accuracy of the short range bridge function procedure. The larger deviations from the Ichimaru expression compared to the deviations from the Rosenfeld expression suggest that the omission of the fourth order Widom coefficient leads to respectable inaccuracies in the OCP screening potential.

The extracted bridge functions $B(x)$ and extracted screening potentials $\ln{[y(x)]}$ in the range $0\leq{x}\leq1.5$, for all the $17$ OCP state points of interest, are illustrated in Fig.\ref{fig:bridge_short_color}. Both the OCP bridge function and OCP screening potential have a smooth predictable pattern with respect to the coupling parameter within the short range and the families of the $B(x;\Gamma)$ as well as the $\ln{[y(x;\Gamma)]}$ curves are again non-intersecting. Taking into account the monotonic polynomial expression for the direct correlation function below its asymptotic limit, the monotonic small argument polynomial expansion of the screening potential and the non-linear closure equation in the short range $B(x)=\ln{[y(x)]}+c(x)+1$, it is no surprise that the short range bridge function has a monotonic polynomial form. It should also be pointed out that, regardless of the coupling parameter, the global extremum of both correlation functions is always obtained at the origin $x=0$, that the slope of the logarithm cavity appears to be zero at the origin (in accordance with the Widom expansion) and that the slope of the bridge function appears to be zero at the origin (although the contact region is somewhat obscured by large local extraction uncertainties).

Finally, the extracted bridge functions $B(x)$ and extracted screening potentials $\ln{[y(x)]}$ within the short range $0\leq{x}\leq1.5$ are compared to the OCP bridge function parametrization of Iyetomi \emph{et al.}\,\cite{bridOCP1} and the OCP screening potential parametrization of Ogata\,\cite{bridOCP2} for four state points in Fig.\ref{fig:error_bars_Ogata} (main, inset). It can be observed that: \textbf{(i)} Regardless of the thermodynamic state, the bridge function and screening potential deviations between the simulation results and the parametrizations always increase towards the origin. \textbf{(ii)} The absolute deviations  between the simulation results and the respective parametrizations are nearly constant for all thermodynamic states. However, owing to the bridge function and screening potential magnitude decrease as the coupling parameter decreases, the absolute relative deviations increase as the coupling parameter decreases. \textbf{(iii)} The statistical errors are much smaller than the deviations from the Iyetomi and Ogata parametrizations. They are rather uniformly spread with the exception of the neighborhood of the origin, where they substantially increase (see the emergence of spiky features and local loss of smoothness). \textbf{(iv)} The statistical errors are nearly independent of the state point, which again implies that their relative value increases as the coupling parameter decreases. A more systematic comparison with the Iyetomi parametrization and a comprehensive presentation of the overall uncertainties can be found in the supplementary material\,\cite{supplem1}.

Tabulated full range OCP bridge functions for the $17$ thermodynamic states of interest have been made freely available online\,\cite{supplem2}. To be more specific, bridge function data are provided for the average as well as the standard deviation of the average in the interval $0.01<x\leq5$ that are discretized by reduced distance steps of $\delta{r}/d=0.01$. Therefore, given the $\Delta{r}/d=0.002$ bin width employed in the histograms for the extraction of the radial distribution function (intermediate \& long range) and for the extraction of the cavity distribution function (short range), the bridge function data have been down-sampled by five. It is pointed out that the starting reduced distance $x$ varies depending on the thermodynamic state, since some close-to-the-origin distances that are judged to be poorly sampled have been removed from the dataset.

\subsection{Extrapolation at the origin}\label{subsec:extrapolation_origin}

\noindent The magnitude of the bridge function at the origin, apart from its relation to the zeroth order Widom coefficient $y_0$, has played a significant role in the development of integral equation theory approximations and in the assessment of their accuracy\,\cite{zeroIET1,zeroIET2}. In addition, there exists a phenomenological freezing criterion that is based on the constant magnitude of $B(0)$ which has been conjectured to be nearly independent of the interaction potential\,\cite{zerosep1}. For these reasons, the magnitude of the OCP bridge function at the origin deserves a meticulous analysis.

Due to the finiteness of the bin width that is employed in the histogram method and owing to the effective bin position that lies at the centre of each bin in the limit of infinitesimal widths, the bridge function is not accessible at distances shorter than $\Delta{r}/2=0.001d$. In practice, our non-accessibility range depends on the coupling parameter and extends up to larger distances, since the shortest distances are not sampled uniformly and, thus, host large statistical errors. In absence of a rigorous small-argument expansion for the bridge function, an extrapolation based on its functional behavior at very short distances should lead to values whose accuracy level would be impossible to evaluate. Here, a simple technique will be followed for the computation of the contact value of the OCP bridge function $B(0)$, which is based on the Widom series representation of the screening potential and on the so-called zero separation theorems\,\cite{zerosep1,zerosep2,zerosep3}. This technique has already been applied to the YOCP\,\cite{Yukawiso}, but it will be briefly repeated in what follows, due to the well-known subtleties that are involved in the inclusion of the uniform charge neutralizing background to the OCP Hamiltonian.

\begin{table*}
	\caption{\emph{1st-6th column:} The extrapolated value of the bridge function at the origin $B_{\mathrm{zsep}}(0)$, as calculated from the zero-separation expression $B_{\mathrm{zsep}}(0)=y_0-\mu_{\mathrm{T}}+2u_{\mathrm{ex}}-\delta+1$, together with all four non-trivial contributions to its magnitude for the $17$ OCP state points of interest. The reduced excess internal energy $u_{\mathrm{ex}}$ and the reduced inverse isothermal compressibility $\mu_{\mathrm{T}}$ are calculated from the OCP equation of state of Ref.\cite{sizecor4}, the zeroth order Widom coefficient $y_0$ is computed by least square fitting input from the specially designed long cavity MD simulations and the integral residual term $\delta$ is computed with input from the accurate standard MD simulations. The contribution of the integral residual term $\delta$ to $B_{\mathrm{zsep}}(0)$ is the smallest for all the $17$ state points, as expected by the fact that the direct correlation function approaches its asymptotic limit already around $x\sim1.6-1.8$ and the fact that the correlation void $g(x)\simeq0$ begins around $x\sim0.8-1.3$, which imply that the integrand factor $g(x)\left[c(x)+\beta{u}(x)\right]$ is non-zero only within a fraction of the first coordination cell $1.05\lesssim{x}\lesssim1.7$. \emph{7th-8th column:} The extrapolated value of the bridge function at the origin $B_{\mathrm{fit}}(0)$, as calculated by least square fitting the monotonic short range bridge function with a polynomial and singling out the constant term, together with the absolute relative deviations between $B_{\mathrm{zsep}}(0)$ and $B_{\mathrm{fit}}(0)$ for the $17$ OCP state points of interest. \emph{9th-10th column:} The extrapolated value of the bridge function at the origin $B_{\mathrm{Iye}}(0)$, as calculated by setting $x=0$ to the Iyetomi \emph{et al.}\,\cite{bridOCP1} parametrization, together with the absolute relative deviations between $B_{\mathrm{zsep}}(0)$ and $B_{\mathrm{Iye}}(0)$ for the $17$ OCP state points of interest.}\label{ZeroSeptable}
	\centering
	\begin{tabular}{cccccc|cccc}\hline
$\Gamma$           & $B_{\mathrm{zsep}}(0)$ & $y_0$                & $\mu_{\mathrm{T}}$    & $u_{\mathrm{ex}}$     & $\delta$               & $B_{\mathrm{fit}}(0)$ & $e_{\mathrm{fit}}$   & $B_{\mathrm{Iye}}(0)$ & $e_{\mathrm{Iye}}$                   \\ \hline
 \,\,\,$170$\,\,\, & \,\,\,$-47.074$\,\,\,  & \,\,\,$182.31$\,\,\, & \,\,\,$-65.858$\,\,\, & \,\,\,$-149.96$\,\,\, & \,\,\,$-3.6870$\,\,\,  & \,\,\,$-46.993$\,\,\, & \,\,\,$0.17\%$\,\,\, & \,\,\,$-44.580$\,\,\, & \,\,\,$5.30\%$\,\,\,                 \\
 \,\,\,$160$\,\,\, & \,\,\,$-44.028$\,\,\,  & \,\,\,$171.64$\,\,\, & \,\,\,$-61.885$\,\,\, & \,\,\,$-141.03$\,\,\, & \,\,\,$-3.5104$\,\,\,  & \,\,\,$-43.969$\,\,\, & \,\,\,$0.13\%$\,\,\, & \,\,\,$-41.274$\,\,\, & \,\,\,$6.25\%$\,\,\,                 \\
 \,\,\,$150$\,\,\, & \,\,\,$-40.961$\,\,\,  & \,\,\,$160.99$\,\,\, & \,\,\,$-57.912$\,\,\, & \,\,\,$-132.11$\,\,\, & \,\,\,$-3.3481$\,\,\,  & \,\,\,$-40.950$\,\,\, & \,\,\,$0.03\%$\,\,\, & \,\,\,$-38.024$\,\,\, & \,\,\,$7.17\%$\,\,\,                 \\
 \,\,\,$140$\,\,\, & \,\,\,$-37.907$\,\,\,  & \,\,\,$150.33$\,\,\, & \,\,\,$-53.940$\,\,\, & \,\,\,$-123.18$\,\,\, & \,\,\,$-3.1804$\,\,\,  & \,\,\,$-37.938$\,\,\, & \,\,\,$0.08\%$\,\,\, & \,\,\,$-34.831$\,\,\, & \,\,\,$8.11\%$\,\,\,                 \\
 \,\,\,$130$\,\,\, & \,\,\,$-34.909$\,\,\,  & \,\,\,$139.66$\,\,\, & \,\,\,$-49.970$\,\,\, & \,\,\,$-114.26$\,\,\, & \,\,\,$-2.9762$\,\,\,  & \,\,\,$-34.933$\,\,\, & \,\,\,$0.07\%$\,\,\, & \,\,\,$-31.699$\,\,\, & \,\,\,$9.20\%$\,\,\,                 \\
 \,\,\,$120$\,\,\, & \,\,\,$-31.904$\,\,\,  & \,\,\,$128.98$\,\,\, & \,\,\,$-46.001$\,\,\, & \,\,\,$-105.34$\,\,\, & \,\,\,$-2.7953$\,\,\,  & \,\,\,$-31.938$\,\,\, & \,\,\,$0.11\%$\,\,\, & \,\,\,$-28.631$\,\,\, & \,\,\,$10.3\%$\,\,\,                 \\
 \,\,\,$110$\,\,\, & \,\,\,$-28.907$\,\,\,  & \,\,\,$118.31$\,\,\, & \,\,\,$-42.034$\,\,\, & \,\,\,$-96.429$\,\,\, & \,\,\,$-2.6055$\,\,\,  & \,\,\,$-28.955$\,\,\, & \,\,\,$0.17\%$\,\,\, & \,\,\,$-25.630$\,\,\, & \,\,\,$11.3\%$\,\,\,                 \\
 \,\,\,$100$\,\,\, & \,\,\,$-25.926$\,\,\,  & \,\,\,$107.64$\,\,\, & \,\,\,$-38.068$\,\,\, & \,\,\,$-87.521$\,\,\, & \,\,\,$-2.4097$\,\,\,  & \,\,\,$-25.987$\,\,\, & \,\,\,$0.23\%$\,\,\, & \,\,\,$-22.702$\,\,\, & \,\,\,$12.4\%$\,\,\,                 \\
 \,\,\,$ 90$\,\,\, & \,\,\,$-22.978$\,\,\,  & \,\,\,$96.948$\,\,\, & \,\,\,$-34.105$\,\,\, & \,\,\,$-78.618$\,\,\, & \,\,\,$-2.2067$\,\,\,  & \,\,\,$-23.036$\,\,\, & \,\,\,$0.25\%$\,\,\, & \,\,\,$-19.850$\,\,\, & \,\,\,$13.6\%$\,\,\,                 \\
 \,\,\,$ 80$\,\,\, & \,\,\,$-20.032$\,\,\,  & \,\,\,$86.263$\,\,\, & \,\,\,$-30.144$\,\,\, & \,\,\,$-69.723$\,\,\, & \,\,\,$-2.0075$\,\,\,  & \,\,\,$-20.107$\,\,\, & \,\,\,$0.38\%$\,\,\, & \,\,\,$-17.081$\,\,\, & \,\,\,$14.7\%$\,\,\,                 \\
 \,\,\,$ 70$\,\,\, & \,\,\,$-17.142$\,\,\,  & \,\,\,$75.552$\,\,\, & \,\,\,$-26.187$\,\,\, & \,\,\,$-60.837$\,\,\, & \,\,\,$-1.7923$\,\,\,  & \,\,\,$-17.206$\,\,\, & \,\,\,$0.38\%$\,\,\, & \,\,\,$-14.400$\,\,\, & \,\,\,$16.0\%$\,\,\,                 \\
 \,\,\,$ 60$\,\,\, & \,\,\,$-14.285$\,\,\,  & \,\,\,$64.838$\,\,\, & \,\,\,$-22.233$\,\,\, & \,\,\,$-51.961$\,\,\, & \,\,\,$-1.5653$\,\,\,  & \,\,\,$-14.341$\,\,\, & \,\,\,$0.39\%$\,\,\, & \,\,\,$-11.817$\,\,\, & \,\,\,$17.3\%$\,\,\,                 \\
 \,\,\,$ 50$\,\,\, & \,\,\,$-11.465$\,\,\,  & \,\,\,$54.119$\,\,\, & \,\,\,$-18.285$\,\,\, & \,\,\,$-43.099$\,\,\, & \,\,\,$-1.3288$\,\,\,  & \,\,\,$-11.523$\,\,\, & \,\,\,$0.50\%$\,\,\, & \,\,\,$-9.3411$\,\,\, & \,\,\,$18.5\%$\,\,\,                 \\
 \,\,\,$ 40$\,\,\, & \,\,\,$-8.7142$\,\,\,  & \,\,\,$43.371$\,\,\, & \,\,\,$-14.344$\,\,\, & \,\,\,$-34.256$\,\,\, & \,\,\,$-1.0821$\,\,\,  & \,\,\,$-8.7676$\,\,\, & \,\,\,$0.61\%$\,\,\, & \,\,\,$-6.9847$\,\,\, & \,\,\,$19.8\%$\,\,\,                 \\
 \,\,\,$ 30$\,\,\, & \,\,\,$-6.0611$\,\,\,  & \,\,\,$32.596$\,\,\, & \,\,\,$-10.412$\,\,\, & \,\,\,$-25.440$\,\,\, & \,\,\,$-0.81086$\,\,\, & \,\,\,$-6.1026$\,\,\, & \,\,\,$0.68\%$\,\,\, & \,\,\,$-4.7641$\,\,\, & \,\,\,$21.4\%$\,\,\,                 \\
 \,\,\,$ 20$\,\,\, & \,\,\,$-3.5479$\,\,\,  & \,\,\,$21.780$\,\,\, & \,\,\,$-6.4981$\,\,\, & \,\,\,$-16.671$\,\,\, & \,\,\,$-0.51539$\,\,\, & \,\,\,$-3.5774$\,\,\, & \,\,\,$0.83\%$\,\,\, & \,\,\,$-2.7009$\,\,\, & \,\,\,$23.9\%$\,\,\,                 \\
 \,\,\,$ 10$\,\,\, & \,\,\,$-1.3188$\,\,\,  & \,\,\,$10.881$\,\,\, & \,\,\,$-2.6193$\,\,\, & \,\,\,$-7.9972$\,\,\, & \,\,\,$-0.17532$\,\,\, & \,\,\,$-1.3090$\,\,\, & \,\,\,$0.75\%$\,\,\, & \,\,\,$-0.8230$\,\,\, & \,\,\,$37.6\%$\,\,\,                 \\ \hline
	\end{tabular}\vspace*{-1.00mm}
\end{table*}

Setting $r=0$ to the OZ equation and to the exact closure equation, employing $g(0)=0$ which is valid for pair interactions that are divergent at the origin, substituting for the logarithm of the cavity distribution function via the Widom series representation $\ln{[y(r)]}|_{r=0}=y_0$ and combining the above, we get the expression
\begin{equation*}
B(0)=y_0-n\int\,c(r)h(r)d^3r\,.
\end{equation*}
We proceed by adding \& subtracting the asymptotic limit of the direct correlation function $-\beta{u}(r)$ within the $c(r)$ factor and utilizing the definition of the OCP reduced excess internal energy $u_{\mathrm{ex}}=(1/2)n\beta\int\,u(r)h(r)d^3r$. Then, we substitute for $h(r)=g(r)-1$ and we employ the statistical relation for the OCP reduced inverse isothermal compressibility $\mu_{\mathrm{T}}=1-n\int\,[c(r)+\beta{u}(r)]d^3r$. Utilizing reduced units and spherical coordinates for the remaining integral $-n\int\,[c(r)+\beta{u}(r)]g(r)d^3r$, we obtain
\begin{equation*}
B(0)=y_0-\mu_{\mathrm{T}}+2u_{\mathrm{ex}}+1-3\int_0^{\infty}x^2g(x)\left[c(x)+\beta{u}(x)\right]dx\,.
\end{equation*}
Introducing the $\delta=3\int_0^{\infty}x^2g(x)\left[c(x)+\beta{u}(x)\right]dx$ compact notation for brevity, where the quantity $\delta$ possesses the smallest value amongst the non-trivial contributions regardless of the coupling parameter, we end up with
\begin{equation*}
B(0)=y_0-\mu_{\mathrm{T}}+2u_{\mathrm{ex}}-\delta+1\,.
\end{equation*}
The four non-trivial contributions to $B(0)$ can be easily determined from readily available input, namely by performing thermodynamic integrations and differentiations of very accurate OCP equations of state ($u_{\mathrm{ex}},\,\mu_{\mathrm{T}}$)\,\cite{sizecor4}, by fitting the Widom series representation of the logarithm of the cavity distribution function to the output of our specially designed long cavity MD simulations ($y_0$) and by employing the extracted radial distribution functions and direct correlation functions from our accurate standard MD simulations ($\delta$). It is noted that tail corrections in the $\delta$ evaluation should be completely negligible given the short range nature of the $c(x)+\beta{u}(x)$ factor of the integrand. It should also be pointed out that the above zero-separation technique formally still constitutes an extrapolation method due to the manner in which $y_0$ is determined. Extrapolations could be completely avoided by using another type of specially designed simulations that are based on the insertion of test particles in order to extract the exact $\ln{[y(r)]}|_{r=0}=y_0$ value at the origin\,\cite{zerosep4}. However, as concluded in section \ref{subsec:extraction_short}, the extrapolated $y_0$ values are accurate enough so that an additional series of simulations is considered to be redundant. This is a fortunate conclusion, since particle insertion methods can become grossly inefficient at strong coupling\,\cite{zerosep5}.

The contact bridge function values that result from our extrapolation method are provided in Table \ref{ZeroSeptable} together with the four non-trivial contributions. In addition, the $B(0)$ values that result from the extrapolation of a polynomial fit of our short range bridge function and the $B(0)$ values that emerge from the Iyetomi \emph{et al.} parametrization are also provided.

\section{The OCP bridge function parametrization}\label{sec:parametrization}

\noindent A transparent strategy has been developed for the acquisition of an accurate analytical representation for the extracted OCP bridge functions. Initially, the oscillatory decaying behavior of the bridge function within the intermediate and long range intervals is fitted with a combination of exponential decays and cosines or sines. Subsequently, the monotonic behavior of the bridge function within the short and intermediate range intervals is fitted with a high order polynomial. Afterwards, as a consequence of the deliberate relatively extended overlap of the above two fitting ranges, the two distinct fitting functions can be smoothly combined into a unique fitting function that remains accurate in the entire range. This is simply accomplished with the aid of a sigmoid switching function that is bounded within $[0,1]$. Finally, this procedure should be repeated for the $17$ OCP state points of interest leading to datasets for the numerous coefficients involved. The OCP bridge function parametrization is completed by seeking for the preferably monotonic fits that capture the coupling parameter dependence of these coefficients.

\subsection{Intermediate and long range parametrization}\label{subsec:parametrization_long}

\noindent The intermediate and long range OCP bridge function is fitted in the interval $1.4\leq{x}\leq3.0$ with the function
\begin{align}
&B_{\mathrm{IL}}(x,\Gamma)=l_0(\Gamma)\exp{\left[-l_1(\Gamma)(x-x_0)-\zeta_1x^2\right]}\times\quad\,\,\,\,\label{longbridgege}\\&\,\,\,\,\,\,\,\,\,\,\left\{\cos{\left[l_2(\Gamma)(x-x_0)\right]}+l_3(\Gamma)\exp{\left[-\zeta_2(x-x_0)\right]}\right\}\,.\nonumber
\end{align}
This fitting expression is inspired by a parametrization that accurately describes the long range bridge functions of dense hard sphere systems\,\cite{bridgHS4}. In the above; the coefficients $x_0,\,\zeta_1,\,\zeta_2$ are constant with $x_0=1.44$, $\zeta_1=0.3$, $\zeta_2=3.5$ (recall the observation that the oscillatory peak \& trough positions of the bridge functions are nearly independent of the coupling parameter), whereas the coefficients $l_0,\,l_1,\,l_2,\,l_3$ are functions of the coupling parameter that are given by
\begin{align}
l_0(\Gamma)&=\Gamma\left\{0.25264-0.31615\ln{(\Gamma)}+0.13135[\ln{(\Gamma)}]^2-\right.\nonumber\\&\,\,\,\,\,\,\left.\quad\,\,\,0.023044[\ln{(\Gamma)}]^3+0.0014666[\ln{(\Gamma)}]^4\right\}\,,\label{longbridgec1}\\
l_1(\Gamma)&=\Gamma^{1/6}\left\{-12.665+20.802\ln{(\Gamma)}-9.6296[\ln{(\Gamma)}]^2+\right.\nonumber\\&\,\,\,\,\,\,\left.\quad\quad\,\,\,\,1.7889[\ln{(\Gamma)}]^3-0.11810[\ln{(\Gamma)}]^4\right\}\,,\label{longbridgec2}\\
l_2(\Gamma)&=\Gamma^{1/6}\left\{15.285-14.076\ln{(\Gamma)}+5.7558[\ln{(\Gamma)}]^2-\right.\nonumber\\&\,\,\,\,\,\,\left.\quad\quad\,\,\,\,1.0188[\ln{(\Gamma)}]^3+0.06551[\ln{(\Gamma)}]^4\right\}\,,\label{longbridgec3}\\
l_3(\Gamma)&=\Gamma^{1/6}\left\{35.330-40.727\ln{(\Gamma)}+16.690[\ln{(\Gamma)}]^2-\right.\nonumber\\&\,\,\,\,\,\,\left.\quad\quad\,\,\,\,2.8905[\ln{(\Gamma)}]^3+0.18243[\ln{(\Gamma)}]^4\right\}\,.\label{longbridgec4}
\end{align}
These fitting expressions are exclusively of the type $l_i(\Gamma)/\Gamma^{s}=\textstyle\sum_{j=0}^{4}l_{i}^{j}(\ln{\Gamma})^{j}$ and inspired by the Iyetomi \emph{et al.} OCP bridge function parametrization\,\cite{bridOCP1}. In general, the presence of $\ln{(\Gamma)}$ factors is typical in the parametrization of different OCP properties such as the excess internal energy\,\cite{IchiRev1,HNCgood1,SMSAminb} or radial distribution function\,\cite{logamma0} and even emerges in exact low $\Gamma-$expansions of the excess internal energy beyond the Debye-H{\"u}ckel term\,\cite{logamma1,logamma2,logamma3}. The mean absolute relative errors of the $l_0(\Gamma)$, $l_1(\Gamma)$, $l_2(\Gamma)$ and $l_3(\Gamma)$ fits are $2.36\%$, $1.98\%$, $0.33\%$ and $3.49\%$, respectively. Note that the mean absolute relative errors of the $B_{\mathrm{IL}}(x,\Gamma)$ fits for each coupling parameter cannot be reliably estimated due to the $B_{\mathrm{IL}}(x,\Gamma)$ zero crossings.

It can be easily proven that the $l_0(\Gamma)$, $l_1(\Gamma)$, $l_2(\Gamma)$ and $l_3(\Gamma)$ fits are monotonic functions of the coupling parameter, which implies that the $B_{\mathrm{IL}}(x,\Gamma)$ parametrization can be interpolated without loss of accuracy and can even be extrapolated with a reasonable loss of accuracy. Since the fitting interval has been contained within $1.4\leq{x}\leq3.0$, the $B_{\mathrm{IL}}(x,\Gamma)$ parametrization can only reliably describe the first oscillation, \emph{i.e.}, the first three zero crossings that contain the first two extrema, but not the secondary oscillations that follow beyond $x=3$. Efforts to extend the $B_{\mathrm{IL}}(x,\Gamma)$ parametrization led to non-monotonic $l_i(\Gamma)$ coefficients and were, thus, discarded. This is most probably a consequence of the statistical errors which, in spite of their small magnitude, might obscure the real bridge function dependence on the coupling parameter at the long range, where its magnitude is naturally very small. Nevertheless, the inclusion of the entire first oscillation in our parametrization constitutes a significant improvement over the Iyetomi parametrization that decays exponentially to zero beyond the first extremum.

\subsection{Intermediate and short range parametrization}\label{subsec:parametrization_short}

\noindent The intermediate and short range bridge function is fitted in the interval $0.2\leq{x}\leq1.8$ with a fifth order polynomial
\begin{align}
B_{\mathrm{IS}}(x,\Gamma)&=s_0(\Gamma)+s_2(\Gamma)x^2+s_3(\Gamma)x^3+\qquad\qquad\qquad\nonumber\\&\,\,\,\,\,\,\,\,s_4(\Gamma)x^4+s_5(\Gamma)x^5\,,\label{shortbridgege}
\end{align}
This fitting expression stems from the short range version of the exact closure equation $B(x)=\ln{[y(x)]}+c(x)+1$, the truncated Widom series representation of the screening potential $\ln{[y(x)]}=y_0+y_2x^2+y_4x^4$ and the analytical SMSA OCP solution for the direct correlation function $c(x)=c_0+c_2x^2+c_3x^3+c_5x^5$\,\cite{SMSAmina}. Taking into account the documented success of the Widom and SMSA expressions, the $B_{\mathrm{IS}}(x,\Gamma)$ parametrization is expected to be very accurate. In fact, the mean absolute relative error of the above fit varies within $0.12-1.99\%$ depending on the coupling parameter and its average value is merely $0.32\%$. It should be noted that the absence of a linear term is consistent with the empirical observation that the bridge function slope is very close to unity at the origin. Alternative polynomial fits which featured a linear term, lacked a cubic term or included higher order terms were attempted that proved to be less accurate. All unknown coefficients $s_0,\,s_2,\,s_3,\,s_4,\,s_5$ have a strong dependence on the coupling parameter that is described by
\begin{align}
s_0(\Gamma)&=\Gamma\left\{0.076912-0.10465\ln{(\Gamma)}+\right.\label{shortbridgec1}\\&\,\,\,\,\,\,\left.\quad\,\,0.0056629[\ln{(\Gamma)}]^2+0.00025656[\ln{(\Gamma)}]^3\right\}\,,\nonumber\\
s_2(\Gamma)&=\Gamma\left\{0.068045-0.036952\ln{(\Gamma)}+\right.\label{shortbridgec2}\\&\,\,\,\,\,\,\left.\quad\,\,0.048818[\ln{(\Gamma)}]^2-0.0048985[\ln{(\Gamma)}]^3\right\}\,,\nonumber\\
s_3(\Gamma)&=\Gamma\left\{-0.30231+0.30457\ln{(\Gamma)}-\right.\label{shortbridgec3}\\&\,\,\,\,\,\,\left.\quad\,\,0.11424[\ln{(\Gamma)}]^2+0.0095993[\ln{(\Gamma)}]^3\right\}\,,\nonumber\\
s_4(\Gamma)&=\Gamma\left\{0.25111-0.26800\ln{(\Gamma)}+\right.\label{shortbridgec4}\\&\,\,\,\,\,\,\left.\quad\,\,0.082268[\ln{(\Gamma)}]^2-0.0064960[\ln{(\Gamma)}]^3\right\}\,,\nonumber\\
s_5(\Gamma)&=\Gamma\left\{-0.061894+0.066811\ln{(\Gamma)}-\right.\label{shortbridgec5}\\&\,\,\,\,\,\,\left.\quad\,\,0.019140[\ln{(\Gamma)}]^2+0.0014743[\ln{(\Gamma)}]^3\right\}\,.\nonumber
\end{align}
These fitting expressions are again exclusively of the type $s_i(\Gamma)/\Gamma=\textstyle\sum_{j=0}^{3}s_{i}^{j}(\ln{\Gamma})^{j}$. The average absolute relative errors of the $s_0(\Gamma)$,\,$s_2(\Gamma)$,\,$s_3(\Gamma)$,\,$s_4(\Gamma)$ and $s_5(\Gamma)$ fits are $0.13\%$, $0.28\%$, $0.70\%$, $4.49\%$ and $3.74\%$, respectively.

It can be proven that the $s_0(\Gamma)$, $s_2(\Gamma)$, $s_3(\Gamma)$, $s_4(\Gamma)$ and $s_5(\Gamma)$ fits are monotonic functions of the coupling parameter, which suggests that also the $B_{\mathrm{IS}}(x,\Gamma)$ parametrization can be interpolated without loss of accuracy and can be extrapolated (even in the supercooled regime) with a reasonable loss of accuracy. It is worth pointing out that the lower endpoint of the fitting interval is selected to be rather high in an effort to avoid the statistical errors that characterize the neighbourhood of the origin, especially for the lowest coupling parameters probed.

\subsection{Full range parametrization}\label{subsec:parametrization_full}

\noindent The intermediate \& long range and intermediate \& short range OCP bridge function parametrizations are unified with the introduction of a switching function in a single parametrization that is strictly valid in the interval ${x}\leq3.0$, but can be employed in the entire range of distances. Overall, we have
\begin{align}
B(x,\Gamma)&=[1-f(x)]B_{\mathrm{IS}}(x,\Gamma)+f(x)B_{\mathrm{IL}}(x,\Gamma)\,,\label{fullbridgege}\\
f(x)&=\frac{1}{2}\left\{1+\mathrm{erf}\left[\zeta_3(x-x_1)\right]\right\}\,,\label{fullbridgesw}
\end{align}
where $x_1$,\,$\zeta_3$ have the constant values $x_1=1.5$, $\zeta_3=5.0$ that are determined by minimizing the deviations of the parametrization from the extracted data and where $f(x)$ denotes the sigmoid switching function that is expressed with the aid of an error function.

\begin{figure}
	\centering
	\includegraphics[width=3.40in]{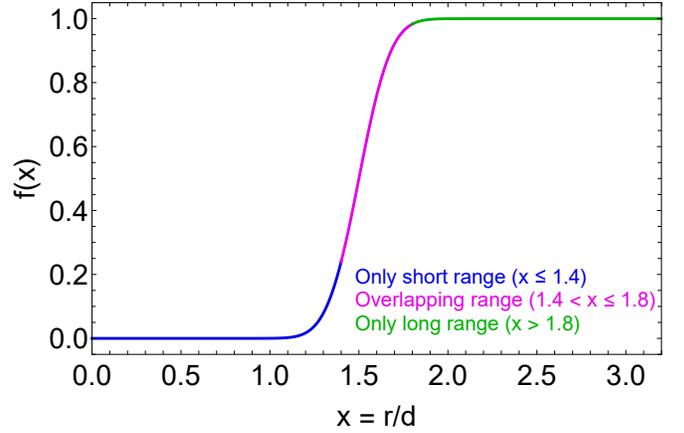}
	\caption{The sigmoid switching function $f(x)$, that is defined in Eq.(\ref{fullbridgesw}), as a function of the reduced distance $x$. The blue line traces values within the fitting interval of the intermediate \& short range alone [$f(x)$ is nearly constant], the green line traces values within the fitting interval of the intermediate \& long range alone [$f(x)$ is constant], the magenta line traces values within the overlap of the two fitting intervals [$f(x)$ is rapidly changing].}\label{fig:switching_function}
\end{figure}

The switching function $f(x)$ is illustrated in Fig.\ref{fig:switching_function}. It is apparent that the largest extent of the interval where this sigmoid function transitions from its lower asymptotic limit of zero to its upper asymptotic limit of unity ($1.3\lesssim{x}\lesssim1.7$) belongs to the overlapping fitting range ($1.4\leq{x}\leq1.8$). This ensures that $B(x,\Gamma)$ properly converges to $B_{\mathrm{IS}}(x,\Gamma)$ at short distances and to $B_{\mathrm{IL}}(x,\Gamma)$ at long distances, but also that $B(x,\Gamma)$ is an accurate representation of the extracted bridge function at intermediate distances.

\begin{figure*}
	\centering
	\includegraphics[width=6.35in]{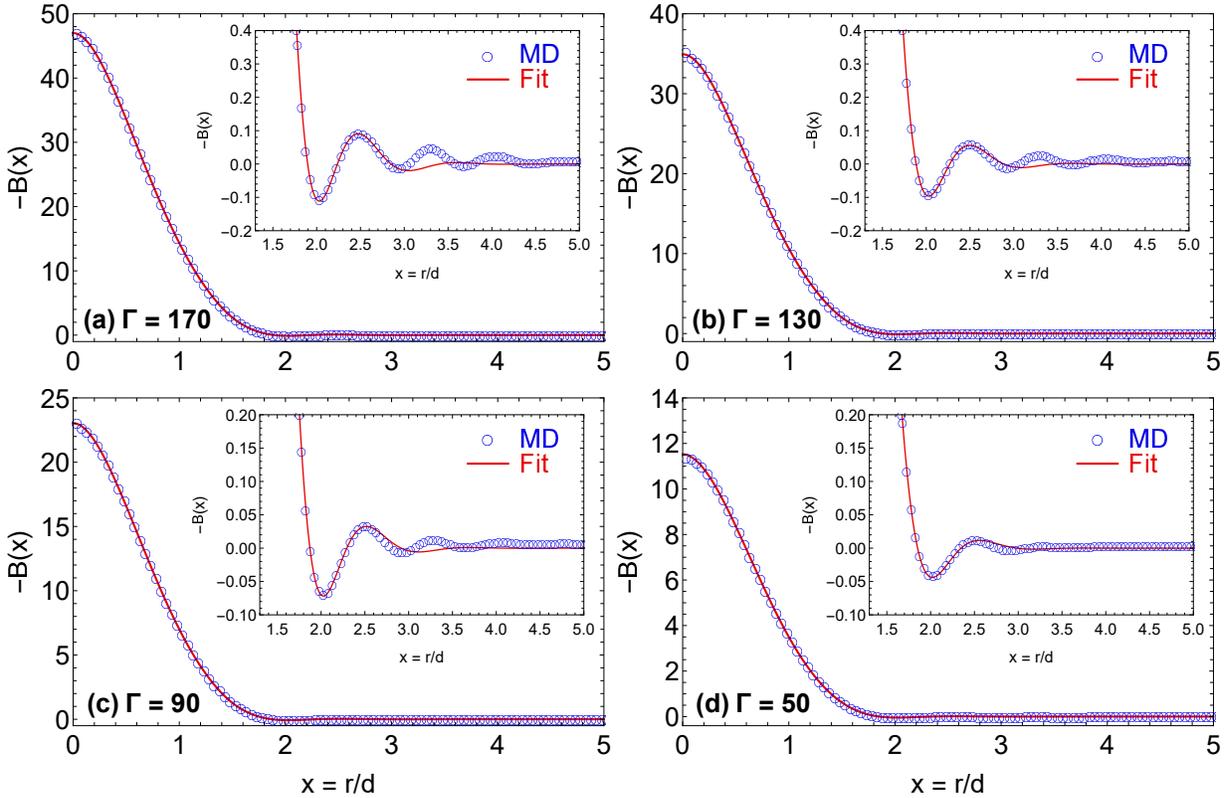}
	\caption{The indirectly extracted OCP bridge functions stemming from combining the OZ inversion method with input from our accurate standard MD simulations \& the cavity distribution method with input from our specially designed long cavity MD simulations (blue circles, downsampled) and as analytically parameterized by the set of Eqs.(\ref{longbridgege},\ref{shortbridgege},\ref{fullbridgege}) (red solid lines). Results for the state points (a)\,$\Gamma=170$,\,(b)\,$\Gamma=130$,\,(c)\,$\Gamma=90$, (d)\,$\Gamma=50$. (Main) The OCP bridge functions in the entire non-trivial interval $0\leq{x}\leq5.0$, where only the monotonic behavior of the short range is discernible. (Inset) Zoom-in on the OCP bridge functions in the intermediate \& long range interval $1.5\leq{x}\leq5.0$, where only the oscillatory decaying behavior is discernible.}\label{fig:bridge_fit}
\end{figure*}

A comparison between the full range parametrization and the indirectly extracted bridge function is featured in Fig.\ref{fig:bridge_fit} for four OCP states. It is confirmed that the sigmoid switching function strategy enables a smooth continuous passage from the monotonic short range behavior to the oscillatory decaying long range behavior. The full range parametrization is characterized by a high level of accuracy, regardless of coupling parameter. The largest deviations are always observed in the near-asymptotic interval $x\geq3.0$ that lies outside the fitting range. Therein, the parameterized OCP bridge function undergoes a very weak stretched oscillation before reaching its zero asymptotic limit, whereas the indirectly extracted OCP bridge function undergoes multiple progressively damped oscillations before reaching its zero asymptotic limit. These two oscillation patterns not only differ in terms of magnitude, but also in terms of wavelength. Taking into account that the neglected secondary oscillations are more prominent close to the crystallization point and that the relative errors in the bridge function extraction are larger close to the Kirkwood point, it can be expected that the parametrization is more accurate for intermediate coupling parameters. Finally, for completeness, we note that the parametrization is deliberately constructed in such a manner so that the bridge function correctly tends to zero everywhere at the weak coupling limit $\Gamma\ll1$. This is straightforward to confirm, since $l_i(\Gamma\to0)=0$ and $s_i(\Gamma\to0)=0$. A more systematic comparison, for all $17$ OCP thermodynamic states of interest, between the indirectly extracted bridge function and the newly-proposed bridge function parametrization can be found in the supplementary material\,\cite{supplem1}.

\section{Application to Yukawa One-Component Plasmas}\label{sec:application}

\noindent It is straightforward that the improved OCP bridge function parametrization can be combined with the OZ equation and the exact non-linear closure equation in an integral equation theory approach that will be remarkably successful for strongly coupled OCP liquids. It is worth emphasizing that empirical approaches based on bridge function parametrizations should be much more accurate than empirical approaches based on direct radial distribution function parametrizations\,\cite{logamma0} in terms of thermodynamic and structural quantities. This is an immediate consequence of the weak sensitivity of the radial distribution function on the bridge function, which implies that small bridge function extraction errors and fitting errors barely propagate to the level of radial distribution functions and thermodynamic quantities. In addition, radial distribution function or structure factor parametrizations are far more complicated than bridge function parameterizations\,\cite{logamma0,logamma4,logamma5,logamma6}. In what follows, it will be shown that the improved OCP bridge function parametrization can constitute the basis of an empirical approach that is remarkably accurate not only for strongly coupled Coulomb liquids, but also for the more extended family of strongly coupled Yukawa liquids.

\subsection{Isomorph-based empirically modified hypernetted chain approach}\label{subsec:IEMHNC_old}

\noindent Yukawa one-component plasmas (YOCP) are model systems that consist of classical point particles which are immersed in a charge neutralizing background. In contrast to the OCP, the uniform background medium is not rigid but polarizable. As a consequence, the charged particles interact via the screened Coulomb (Yukawa) pair potential $u(r)=(Q^2/r)\exp(-r/\lambda)$ where $\lambda$ is the linear screening length. The thermodynamic state points of the YOCP are specified in terms of two independent dimensionless variables, the standard coupling parameter $\Gamma=(\beta{Q^2})/{d}$ and the screening parameter $\kappa={d}/{\lambda}$\,\cite{dustrev1,dustrev2,dustrev3,dustrev4}. In the rigid background limit $\lambda\to\infty$ or $\kappa\to0$, the Yukawa potential collapses to the bare Coulomb potential and the YOCP collapses to the OCP. The YOCP is an important model system of statistical mechanics, since it explores the full range of potential softness from the long range Coulomb interactions of the OCP for $\kappa=0$ to the ultra-short range interactions of the hard sphere system for $\kappa\to\infty$. The YOCP also has important practical implications due to its relevance to complex plasmas\,\cite{dustrev3,dustrev4}, colloidal suspensions\,\cite{dustrev5,dustrev6}, ultra-cold neutral plasmas\,\cite{dustrev7,dustrev8} and warm dense matter\,\cite{dustrev9,dustrev0}.

It has recently been demonstrated that strongly coupled YOCP liquids belong to the broad class of R-simple systems\,\cite{isoYOCPg}. Such systems are the subject of isomorph theory\,\cite{isomssm1,isogene1,isogene2,isogene3} and their rigorous definition concerns the relation    $U(\boldsymbol{R}_{\mathrm{a}})<U(\boldsymbol{R}_{\mathrm{b}})\Rightarrow{U}(\mu\boldsymbol{R}_{\mathrm{a}})<U(\mu\boldsymbol{R}_{\mathrm{b}})$, where $U(\boldsymbol{R})$ is the total potential energy, $\boldsymbol{R}=(\boldsymbol{r}_1,...,\boldsymbol{r}_N)$ is the particle configuration, $\boldsymbol{R}_{\mathrm{a}}$ \& $\boldsymbol{R}_{\mathrm{b}}$ are two equal density configurations and $\mu$ denotes a positive constant\,\cite{isogene4}. The definition states that the ordering of the total potential energies of two configurations that are consistent with the same density is maintained when both these two configurations are uniformly scaled to the same different density. It is exact only for systems whose constituents interact via Euler-homogeneous potentials (plus constant), such as inverse power law systems. For all other R-simple systems, the definition should be understood to concern the most physically relevant configurations\,\cite{isogene4}, which reflects the approximate nature of the isomorph theory and its consequences. The practical identification of R-simple systems concerns the existence of strong correlations between their virial $(W)$ and potential energy $(U)$ constant-volume thermal equilibrium fluctuations\,\cite{isogene5}. The Pearson coefficient $R_{\mathrm{WU}}$ is employed to quantify the strength of $W-U$ correlations, with $R_{WU}\gtrsim0.9$ corresponding to R-simple systems; a criterion that allows for a straightforward characterization from canonical (NVT) computer simulations.

R-simple systems possess isomorphic lines or just isomorphs; \emph{i.e.} phase diagram curves of constant excess entropy, along which a large set of structural and dynamic properties are approximately invariant when expressed in properly reduced units where the length is normalized to the mean-cubic inter-particle distance $\Delta=n^{-1/3}$, the energy is normalized to the thermal energy $1/\beta$ and the time is normalized to $n^{-1/3}\sqrt{\beta{m}}$\,\cite{isomssm1,isogene1,isogene2,isogene3}. The invariant properties include the radial distribution function\,\cite{isomssm1} (but not the direct correlation function), the incoherent intermediate scattering function\,\cite{isogene3}, the transport coefficients\,\cite{isogene6} \& several high order structure measures\,\cite{isogene7}. Very recently, a systematic computational investigation revealed that the bridge function belongs to this list of approximately isomorph invariant quantities\,\cite{Yukawiso}.

The isomorph-based empirically modified hypernetted chain (IEMHNC) approximation is a novel integral equation theory approach that is based on the isomorph invariance of the reduced unit bridge functions of R-simple systems\,\cite{ourwork1}. This approximate invariance property leads to a self-consistent closure to the exact relations of integral equation theory provided that two external inputs are available: the closed-form description of the isomorphic curves on the thermodynamic phase diagram and a bridge function parametrization valid along any phase diagram line that has a unique intersection point with any isomorphic curve\,\cite{ourwork1}. Thus, the available isentropic correspondence can map the bridge function from the initial thermodynamic line to the entire phase diagram.

The YOCP exhibits exceptionally strong $W-U$ correlations ($R_{WU}>0.99$) for a quite extended part of the fluid phase covering the entire dense liquid region of the phase diagram\,\cite{Yukawiso,isoYOCPg}. Different methods have been employed to trace out the YOCP isomorphic lines that correspond to different reduced excess entropies and it has been demonstrated that all YOCP isomorphs can be accurately parameterized by\,\cite{Yukawiso,isoYOCPg}
\begin{equation}
\Gamma_{\mathrm{iso}}(\Gamma,\kappa)=\Gamma{e}^{-\alpha\kappa}\left[1+\alpha\kappa+(1/2)(\alpha\kappa)^2\right]=\mathrm{const}\label{YOCPisomorph}
\end{equation}
where $\alpha=\Delta/d=(4\pi/3)^{1/3}$ denotes the ratio of the cubic mean inter-particle distance $\Delta$ over the Wigner-Seitz radius $d$. We note that this closed-form expression was originally proposed as an empirical description of the YOCP crystallization line as determined by MD simulations\,\cite{isogene8,isogene9}. This is consistent with the finding that, to the first order, the isomorphic lines are nearly parallel to the crystallization line\,\cite{isogene0,isogeneL}. Given the approximate isomorph invariance property and the configurational adiabat mapping of Eq.(\ref{YOCPisomorph}), the OCP bridge functions constitute the basis for the construction of YOCP bridge functions via
\begin{equation}
B_{\mathrm{YOCP}}(x,\Gamma,\kappa)=B_{\mathrm{OCP}}[x,\Gamma_{\mathrm{ISO}}(\Gamma,\kappa)]\,.\label{YOCPbridgefunction}
\end{equation}
Overall, the IEMHNC approach for YOCP liquids consists of Eqs.(\ref{OZequation},\ref{OZclosure}), Eqs.(\ref{YOCPisomorph},\ref{YOCPbridgefunction}) and the OCP bridge function parametrization.

The IEMHNC approach has been successfully applied to dense stable Yukawa liquids\,\cite{ourVMHNC,ourwork1,ourwork2}, metastable Yukawa liquids\,\cite{ourwork3} and dense bi-Yukawa liquids\,\cite{ourwork4} employing the Iyetomi \emph{et al.} parametrization of the OCP bridge function\,\cite{bridOCP1}. It should be pointed out that these investigations proceeded the computational demonstration that YOCP bridge functions are approximately isomorph invariant quantities\,\cite{Yukawiso}, thus the bridge function isomorph invariance is referred to as an ansatz therein. A comprehensive benchmarking with available computer simulations of dense YOCP liquids has revealed that this version of the IEMHNC approach has a remarkable accuracy with predictions of structural properties within $2\%$ inside the first coordination cell and predictions of thermodynamic properties within $0.5\%$\,\cite{ourVMHNC,ourwork1}. In addition, a systematic comparison with different advanced integral equation theory approximations has demonstrated that the performance of the original version of the IEMHNC approach is comparable to that of the VMHNC approach \cite{rosVMHNC} but with a $10-80$ times less computational cost depending on the state point\,\cite{ourVMHNC}. In section \ref{subsec:IEMHNC_new}, we shall explore the performance of a new version of the IEMHNC approach for dense YOCP liquids, that employs our improved OCP bridge function parametrization as a building block instead of the Iyetomi parametrization. A possible improved agreement with computer simulations of dense YOCP liquids would serve as additional evidence for the superiority of our parametrization.

\begin{figure*}
	\centering
	\includegraphics[width=6.00in]{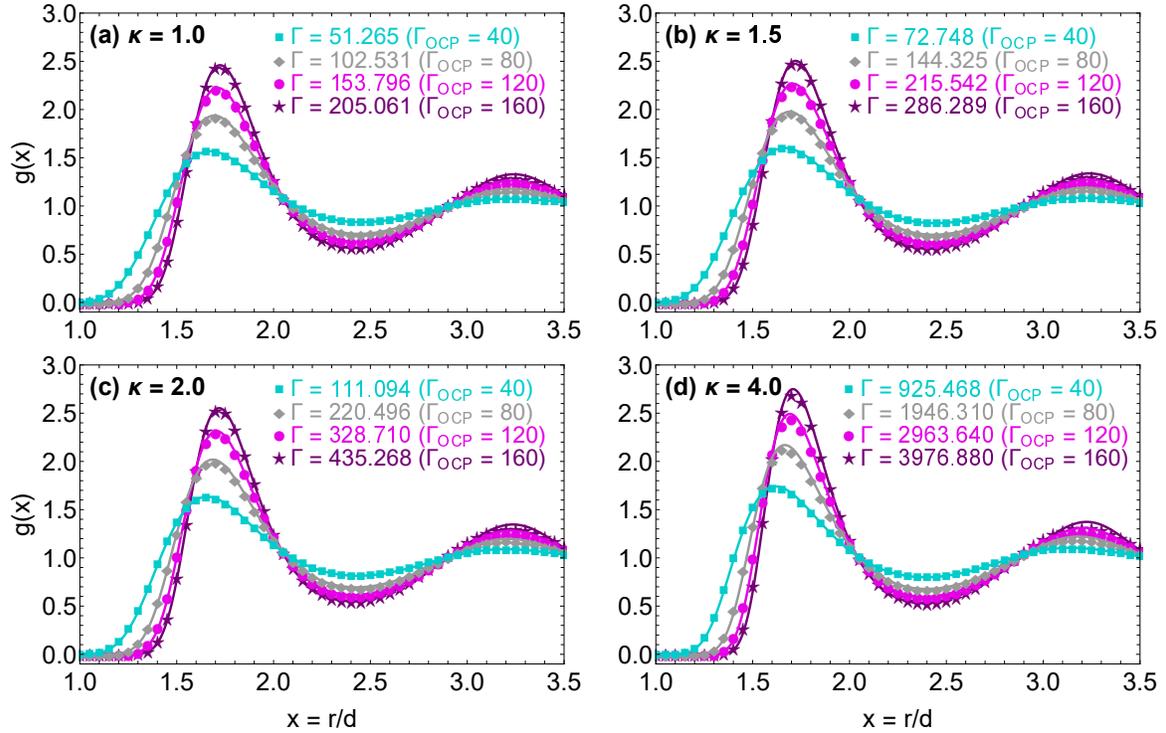}
	\caption{The YOCP radial distribution functions resulting from MD simulations (discrete symbols) and from the new version of the IEMHNC integral equation theory approach that utilizes our improved OCP bridge function parametrization (solid lines). Results for four screening parameters $\kappa=1.0$ (a), $\kappa=1.5$ (b), $\kappa=2.0$ (c), $\kappa=4.0$ (d) and for four coupling parameters whose isomorphic OCP state points have $\Gamma_{\mathrm{OCP}}=160$ (purple, stars), $\Gamma_{\mathrm{OCP}}=120$ (magenta, circles), $\Gamma_{\mathrm{OCP}}=80$ (light gray, diamonds), $\Gamma_{\mathrm{OCP}}=40$ (cyan, squares). The IEMHNC results are nearly indistinguishable from the MD simulations. Note that the MD results were down-sampled in order to improve visibility.}\label{fig:IEMHNC_YOCP}
\end{figure*}

\subsection{Structural and thermodynamic properties}\label{subsec:IEMHNC_new}

\noindent The new version of the IEMHNC approach for YOCP liquids consists of the OZ equation and the non-linear closure equation, see Eqs.(\ref{OZequation},\ref{OZclosure}), of the YOCP isomorph mapping and the approximate reduced-unit isomorph invariance property of bridge functions, see Eqs.(\ref{YOCPisomorph},\ref{YOCPbridgefunction}), as well as of our improved OCP bridge function parametrization, see Eqs.(\ref{longbridgege},\ref{shortbridgege},\ref{fullbridgege}). The numerical solution of this set of equations is achieved with a well-established algorithm that is based on Picard iterations in Fourier space\,\cite{ourVMHNC,ourwork1}.

A graphical comparison between the radial distribution functions resulting from the new IEMHNC approach with the \enquote{exact} radial distribution functions extracted from MD simulations is illustrated in Fig.\ref{fig:IEMHNC_YOCP} for $16$ different YOCP state points (four different screening parameters). The IEMHNC results are nearly indistinguishable from the MD simulations within the first coordination cell, especially for $\kappa=1.0,\,1.5,\,2.0$. Deviations are discernible for $\kappa=4.0$ and the two larger coupling parameters investigated; most probably a consequence of the isomorph invariance property which becomes more approximate as the interaction softness decreases. In addition, small deviations are visible in the neighbourhood of the second maximum of the radial distribution function only for the largest coupling parameter probed, regardless of screening length. This is most probably a consequence of the OCP bridge function parametrization, whose fitting range did not include the second coordination cell.

The MD simulations, the original IEMHNC approach and the updated IEMHNC approach lead to radial distribution functions which are so close, that a simple graphical inspection cannot lead to any safe conclusion regarding which version of the IEMHNC approach is superior. For quantitative comparison, we shall resort to some key functional properties of the radial distribution function that have been extracted in a systematic Langevin Dynamics investigation of the weakly screened YOCP\,\cite{OttLDOCP}. This simulation study probed three values of the screening parameter ($\kappa=0,\,1,\,2$) and numerous values of the coupling parameter between unity and the crystallization value. The functional characteristics extracted from these simulations include the edge of the correlation void [estimated as the position where $g(r)=0.5$], the magnitude as well as the position of the first maximum, the first non-zero minimum, and the second maximum. Extensive tabulations of the radial distribution function characteristics as obtained from these simulations as well as the original and updated versions of the IEMHNC approach are provided in the supplementary material\,\cite{supplem1}.

The main conclusions drawn from this comparison are the following. \emph{In the OCP case}, the updated IEMHNC approach strongly improves the predictions of the original IEMHNC approach regarding the edge of the correlation void, the magnitude \& position of the first maximum and the magnitude \& position of the first non-zero minimum. On the other hand, the second maximum predictions of the two IEMHNC versions are comparable, with the updated version being more accurate for the position and the original version more accurate for the magnitude. Note also that the updated IEMHNC version yields more accurate radial distribution functions not only on average but also distinctly for all $31$ OCP states investigated. The above strongly confirm the superiority of our OCP bridge function parametrization over the Iyetomi parametrization. \emph{In the YOCP case}, regardless of screening parameter, the updated IEMHNC approach again greatly improves the original IEMHNC approach predictions for the edge of the correlation void as well as for the magnitude \& position of the first maximum and first non-zero minimum. Moreover, the IEMHNC accuracy level is observed to be nearly constant for any screening parameter, independently confirming the isomorph invariance property of bridge functions. Overall, the radial distribution functions as computed by the updated IEMHNC version are characterized by a truly unprecedent agreement with the \enquote{exact} radial distribution functions as extracted from computer simulations. The average accuracy level is estimated to be $\sim0.2\%$ within the first coordination cell.

Finally, extensive tabulations of the reduced excess internal energy as obtained from MD simulations available in the literature\,\cite{sizecor4,sizecor5} as well as from the original and updated versions of the IEMHNC approach are provided in the supplementary material\,\cite{supplem1}. The updated version again leads to more accurate predictions regardless of the screening, although it is acknowledged that the improvement over the original version is now less impressive.

Overall, given that the original IEMHNC version has an accuracy comparable to that of the VMHNC approach \cite{ourVMHNC}, it can be safely concluded that the updated version of the IEMHNC approach constitutes by far the most accurate theoretical approach ever devised for predictions of the YOCP structural and thermodynamic properties.

\section{Summary and conclusions}\label{outro}

\noindent The bridge functions of strongly coupled one-component plasmas were systematically computed for $17$ thermodynamic state points uniformly spread from the Kirkwood point to the bcc crystallization point. The intermediate and long range bridge functions were made accessible after application of the Ornstein-Zernike inversion method with structural input from accurate standard canonical (NVT) molecular dynamics simulations, while the short range bridge functions were made accessible after application of the cavity distribution method with structural input from long specially designed canonical molecular dynamics simulations that feature a tagged particle pair. The reliable extraction beyond the correlation void was achieved by utilizing a near-optimal bin width that minimizes grid errors with respect to statistical errors, applying the simplified Lebowitz-Percus expression to correct for the finite size errors, developing a Pad{\'e} approximant strategy to enforce the compressibility sum rule on the structure factors and using Ng's long range decomposition technique to compute real space direct correlation functions. The reliable extraction in the correlation void was achieved by splitting the short range interval into four overlapping windows via the inclusion of a windowing component in the tagged potential and carefully optimizing the biasing component of the tagged potential so that uniform statistics are obtained. Reliable extrapolation at the origin was realized by applying a methodology reminiscent of the so-called zero-separation theorems. A detailed error propagation analysis led to accurate quantification of the uncertainties that are dominated by statistical errors, which are only prominent near the origin.

The one-component plasma bridge functions were parameterized by adopting a switching function strategy. In the intermediate and long range intervals, the oscillatory decaying behavior was fitted with a combination of exponential decays and cosines, as inspired by an existing hard sphere bridge function parametrization. In the short and intermediate range intervals, the monotonic behavior was fitted with a fifth-order polynomial without a linear term, as dictated by the Widom small argument expansion for the logarithm of the cavity distribution function and the solution of the soft mean spherical approximation for the direct correlation function. Owing to the deliberately extended overlap of the two fitting ranges, the two distinct fitting functions were smoothly combined into a unique continuous fitting function valid in the entire range with the aid of a sigmoid switching function that was bounded within zero and unity. All the emerging coupling parameter dependent coefficients were fitted with monotonic functions of the form $u_i(\Gamma)/\Gamma^{s}=\textstyle\sum{u}_{i}^{j}(\ln{\Gamma})^{j}$; a form inspired by established parametrizations of numerous one-component plasma properties. The monotonicity ensures that the parametrization can be state interpolated without any loss of accuracy and also suggests that it can even be state extrapolated with a reasonable loss of accuracy. Overall, the proposed parametrization near-perfectly describes the one-component plasma bridge functions from the contact up to the end of the first coordination cell. It is superior to the existing Iyetomi \emph{et al.} parametrization that suffers from grid errors, decays exponentially to zero beyond the first extremum and cannot be reliably interpolated being based on solely $4$ state point extractions.

We emphasize that the present bridge function extraction and parametrization activities constitute the most comprehensive to be reported in the long history of bridge functions. Thus, one-component plasma bridge functions can be considered to be characterized better than those of Lennard-Jones and even hard sphere systems.

The new parametrization of the one-component plasma bridge function was also employed as a building block of the isomorph-based empirically modified hypernetted chain approximation that was systematically applied to strongly coupled Yukawa one-component plasma liquids. This novel integral equation theory approach is based upon the recently revealed approximate isomorph invariance of the reduced-unit bridge functions of R-simple systems and utilizes an isentropic mapping in order to accurately map the one-component plasma bridge functions to Yukawa one-component plasma bridge functions. An exhaustive comparison with \enquote{exact} results from readily available computer simulations revealed that this approach constitutes by far the most accurate theoretical method ever devised for predictions of the structural and thermodynamic properties of Yukawa liquids. Given also its simplicity and its low computational cost, we expect this approach to replace all known integral equation theory approaches that have been devised for Coulomb and for Yukawa liquids. Taking into account the availability of accurate equations of state from computer simulations, this conclusion mainly concerns the calculation of static properties that, however, also constitute external input for the theoretical description of dynamic and transport properties, see for instance the quasi-localized charge approximation for collective modes\,\cite{outrore0,outrore1}, the viscoelastic dynamic density functional theory for wave dispersion\,\cite{outrore2,outrore3}, the moment approach of dynamic density-density correlations\,\cite{outrore4,outrore5}, the mode coupling theory of the glass transition\,\cite{ourwork3,outrore6,outrore7}, the static local field corrected model for the thermal conductivity\,\cite{outrore8,outrore9}.

It is worth mentioning that the new parametrization of the one-component plasma bridge function has already been embedded in a novel dielectric formalism scheme for quantum one-component plasma liquids\,\cite{unpubli1,unpubli2}. This scheme treats quantum effects on the random phase approximation level and correlation effects in an exact classical manner. Systematic comparison with \emph{ab initio} path integral Monte Carlo simulations of paramagnetic electron liquids revealed an unprecedented agreement in terms of both thermodynamic and structural properties.

Future work will address the extraction of the bridge functions of two-dimensional one-component plasmas\,\cite{OCPrevBH,outror10} and of binary ionic mixtures\,\cite{OCPrevBH,outror11} as well as the formulation of the isomorph-based empirically modified hypernetted chain approach for two-dimensional Yukawa one-component plasmas\,\cite{outror12,outror13} and for binary Yukawa mixtures\,\cite{outror14,outror15}. Our bridge function studies have progressed enough so that such extensions can be deemed as rather straightforward albeit still very cumbersome.

\section*{Acknowledgments}

\noindent The authors would like to acknowledge the financial support of the Swedish National Space Agency under grant no.\,143/16. Molecular dynamics simulations were carried out on resources provided by the Swedish National Infrastructure for Computing (SNIC) at the NSC (Link{\"o}ping University) that is partially funded by the Swedish Research Council through grant agreement no.\,2018-05973.

\renewcommand{\theequation}{A\arabic{equation}}
\setcounter{equation}{0}
\section*{APPENDIX: The Iyetomi \emph{et al.} bridge function and the Ogata screening potential parametrizations}

\noindent Iyetomi \emph{et al.} indirectly extracted the OCP bridge functions at four state points ($\Gamma=10,\,40,\,80,\,160$), employing structural input from accurate standard Monte Carlo (MC) simulations\,\cite{bridOCP1}. The long range nature of Coulomb interactions was handled with the traditional Ewald technique, the simulated particle number was $N=1024$, the cubic simulation box length was $L/d=16.2$, the number of configurations was $7\times10^{6}$ and the histogram bin width was $\Delta{r}/d=0.04$.

The intermediate range bridge functions were obtained with the OZ inversion method. The long range extrapolations were based on the enforcement of the compressibility sum rule for the structure factor based on a Pad\'e approximant strategy, while the short range extrapolations were based on the cavity distribution function continuity, the Widom series representation, the Jancovici exact result for the second order Widom coefficient and the assertion that the fourth order Widom coefficient can be simply omitted beyond the lower endpoint of the reliable bridge function data. It is important to point out that the second order truncation of the Widom series was strongly criticized by Rosenfeld\,\cite{rosepyc1,rosepyc2}, who considered the fourth order truncation to be more appropriate.

The \emph{Iyetomi parametrization of the OCP bridge function} ultimately reads as\,\cite{bridOCP1}
\begin{align*}
&B_{\mathrm{OCP}}(r,\Gamma)=\Gamma\left[-b_0(\Gamma)+c_1(\Gamma)\left(\frac{r}{d}\right)^4+c_2(\Gamma)\left(\frac{r}{d}\right)^6\right.\nonumber\\&\,\qquad\qquad\qquad\left.+c_3(\Gamma)\left(\frac{r}{d}\right)^8\right]\exp{\left[-\frac{b_1(\Gamma)}{b_0(\Gamma)}\left(\frac{r}{d}\right)^2\right]}\,,\\
&b_0(\Gamma)=0.258-0.0612\ln{\Gamma}+0.0123(\ln{\Gamma})^2-\frac{1}{\Gamma}\,,\\
&b_1(\Gamma)=0.0269+0.0318\ln{\Gamma}+0.00814(\ln{\Gamma})^2\,,\\
&c_1(\Gamma)=0.498-0.280\ln{\Gamma}+0.0294(\ln{\Gamma})^2\,,\\
&c_2(\Gamma)=-0.412+0.219\ln{\Gamma}-0.0251(\ln{\Gamma})^2\,,\\
&c_3(\Gamma)=0.0988-0.0534\ln{\Gamma}+0.00682(\ln{\Gamma})^2\,.
\end{align*}
Notice that the five coupling parameter dependent coefficients involve three fitting constants, even though computed bridge functions were available for four OCP coupling parameters. This observation implies that extrapolations and even interpolations might lead to a substantial loss of accuracy. The validity range of the Iyetomi \emph{et al.} parametrization was specified to be $5<\Gamma\leq180$. The upper validity threshold of $\Gamma_{\mathrm{th}}^{\mathrm{u}}=180$ simply corresponds to the coupling parameter at the bcc crystallization point, as estimated by early computer simulations. The lower validity threshold of $\Gamma_{\mathrm{th}}^{\mathrm{l}}\simeq5$ is imposed mathematically by the sign switching of the $b_0(\Gamma)$ coefficient. To be more specific, close to $\Gamma\simeq5.25$, the coefficient $b_0$ becomes negative but the coefficient $b_1$ remains positive, which leads to an exponential blow-up of the bridge function at large distances. Notice also that the assumed fitting function form can only  describe the monotonic behavior and the first extremum of the bridge function and that it features an artificial strong exponential decay in the place of the physical oscillatory decay.

Ogata indirectly extracted the logarithm of the OCP cavity distribution function or the OCP screening potential within the range $0\leq{x}\leq1.7$ at the same four OCP state points ($\Gamma=10,\,40,\,80,\,160$), employing structural input from long specially designed cavity MC simulations featuring a tagged particle pair\,\cite{bridOCP2}. The simulated particle number was $N=1000$, consisting of $998$ standard particles and $2$ tagged particles, which leads to a $L/d=16$ box length. The number of configurations was $\sim10^{8}$ and the histogram bin width was $\Delta{r}/d=0.04$. The particle to be subject to a trial displacement in the Metropolis algorithm was chosen with a probability of $25\%$ to be a tagged particle and with a probability of $0.05\%$ to be a standard particle.

The \emph{Ogata parametrization of the OCP screening potential} ultimately reads as\,\cite{bridOCP2}
\begin{equation*}
\frac{\beta{H}_{\mathrm{O}}(x)}{\Gamma} =
 \begin{cases}
 A_1-\displaystyle\frac{1}{4}x^2\left[1-\exp{\left(-\frac{A_2}{x}\right)}\right]\,,x\leq{A}_3 \\
 A_4-A_5x+\displaystyle\frac{\exp{(A_6\sqrt{x}-A_7)}}{x}\,,A_3<x\leq2
 \end{cases}\,\,
\end{equation*}
and extends from the entire short range up to the intermediate range. From the asymptotic limit of the bracketed exponential factor of the short range branch, it becomes evident that the parametrization has been constructed to abide by the Widom small argument expansion and by the Jancovici exact result for the second order Widom OCP coefficient. The seven coupling parameter dependent coefficients are given by
\begin{align*}
&A_1(\Gamma)=1.132-0.0094\ln{\Gamma}\,,\\
&A_2(\Gamma)=2.55-0.043\ln{\Gamma}\,,\\
&A_3(\Gamma)=1.22-0.047\ln{\Gamma}\,,\\
&A_4(\Gamma)=1.356-0.0213\ln{\Gamma}\,,\\
&A_5(\Gamma)=0.456-0.013\ln{\Gamma}\,,\\
&A_6(\Gamma)=9.29+0.79\ln{\Gamma}\,.\\
&A_7(\Gamma)=14.83+1.31\ln{\Gamma}\,.
\end{align*}
Notice that the seven coupling parameter dependent coefficients involve two fitting constants, even though computed screening potentials were available only for four OCP coupling parameters. This observation clearly suggests that extrapolations and even interpolations might lead to a substantial loss of accuracy. Note also that the zeroth order Widom OCP coefficient, which - as aforementioned - essentially controls the enhancement factor of the pycnonuclear reactions in dense astrophysical plasmas owing to charge screening effects, is simply given by $y_0(\Gamma)=A_1(\Gamma)$.

\end{document}